\documentclass[12pt, draftclsnofoot, onecolumn]{IEEEtran}
\usepackage{cite,graphicx,amsmath,amssymb}
\usepackage{subfigure}
\usepackage{citesort}
\usepackage{fancyhdr}
\usepackage{mdwmath}
\usepackage{mdwtab}
\usepackage{balance}
\usepackage{xcolor}
\usepackage{bm}
\usepackage{cite,graphicx,amsmath,amssymb}
\usepackage{subfigure}
\usepackage{citesort}
\usepackage{fancyhdr}
\usepackage{mdwmath}
\usepackage{mdwtab}
\usepackage{balance}
\usepackage{xcolor}
\usepackage{bm}
\usepackage{amssymb}
\usepackage{amsmath}
\usepackage{cite}
\usepackage{url}
\usepackage{xcolor}
\usepackage{cite,graphicx,amsmath,amssymb}
\usepackage{subfigure}
\usepackage{citesort}
\usepackage{fancyhdr}
\usepackage{mdwmath}
\usepackage{mdwtab}
\usepackage{float}
\usepackage{caption}
\usepackage{amsthm}
\usepackage{graphicx}
\usepackage{algorithm}
\usepackage{algorithmic}
\usepackage{multirow}
\usepackage{flafter}
\usepackage{graphicx}
\usepackage{subfigure}
\usepackage{algorithm}
\usepackage{setspace}
\usepackage{algorithmic}
\usepackage{caption}
\usepackage{amssymb}
\usepackage{amsmath}
\usepackage{cite}
\usepackage{url}
\usepackage{xcolor}
\usepackage{cite,graphicx,amsmath,amssymb}
\usepackage{subfigure}
\usepackage{citesort}
\usepackage{fancyhdr}
\usepackage{mdwmath}
\usepackage{mdwtab}
\usepackage{caption}
\usepackage{amsthm}
\usepackage{algorithm}
\usepackage{algorithmic}

\usepackage{multirow}
\usepackage{flafter}
\usepackage{graphicx}
\usepackage{subfigure}
\usepackage{algorithm}
\usepackage{setspace}
\usepackage{algorithmic}
\usepackage{caption}
\newtheorem{remark}{Remark}

\usepackage{amsmath,amsthm,amssymb}

\newtheorem{theorem}{Theorem}

\begin{document}

\title{Machine Learning Empowered Trajectory and Passive Beamforming Design in UAV-RIS Wireless Networks}

\author{
Xiao~Liu,~\IEEEmembership{Student Member,~IEEE,}
Yuanwei~Liu,~\IEEEmembership{Senior Member,~IEEE,}\\
Yue~Chen,~\IEEEmembership{Senior Member,~IEEE,}

\thanks{

X. Liu, Y. Liu, and Y. Chen are with the School of Electronic Engineering and Computer Science, Queen Mary University of London, London E1 4NS, UK. (email: x.liu@qmul.ac.uk; yuanwei.liu@qmul.ac.uk; yue.chen@qmul.ac.uk).

}
}

\maketitle

\vspace{-1.5cm}
\begin{abstract}
A novel framework is proposed for integrating reconfigurable intelligent surfaces (RIS) in unmanned aerial vehicle (UAV) enabled wireless networks, where an RIS is deployed for enhancing the service quality of the UAV. Non-orthogonal multiple access (NOMA) technique is invoked to further improve the spectrum efficiency of the network, while mobile users (MUs) are considered as roaming continuously. The energy consumption minimizing problem is formulated by jointly designing the movement of the UAV, phase shifts of the RIS, power allocation policy from the UAV to MUs, as well as determining the dynamic decoding order. A decaying deep Q-network (D-DQN) based algorithm is proposed for tackling this pertinent problem. In the proposed D-DQN based algorithm, the central controller is selected as an agent for periodically observing the state of UAV-enabled wireless network and for carrying out actions to adapt to the dynamic environment. In contrast to the conventional DQN algorithm, the decaying learning rate is leveraged in the proposed D-DQN based algorithm for attaining a tradeoff between accelerating training speed and converging to the local optimal. Numerical results demonstrate that: 1) In contrast to the conventional Q-learning algorithm, which cannot converge when being adopted for solving the formulated problem, the proposed D-DQN based algorithm is capable of converging with minor constraints; 2) The energy dissipation of the UAV can be significantly reduced by integrating RISs in UAV-enabled wireless networks; 3) By designing the dynamic decoding order and power allocation policy, the RIS-NOMA case consumes 11.7\% less energy than the RIS-OMA case.
\end{abstract}


\section{Introduction}

The fifth generation of wireless networks (5G) is being developed to provide uninterrupted and ubiquitous connectivity to everyone and everything, everywhere, which impose enormous challenges on conventional terrestrial cellular networks. The massive multiple-input-multiple-output (MIMO) technology, which equips base stations (BSs) with an array of active antennas, has been proved to improve the spectrum efficiency of next-generation systems. A related concept, namely, reconfigurable intelligent surfaces (RISs)~\cite{liu2020Reconfigurable}, also referred as intelligent reflecting surfaces (IRSs)~\cite{Qingqing2020Towards} or large intelligent surfaces (LISs)~\cite{najafi2020physics}\footnote[1]{Without loss of generality, we use the name of RIS in the remainder of this paper.}, comprise an array of reflecting elements for reconfiguring the incident signals. Thus, RISs can be regarded as the pathway towards massive MIMO 2.0. RISs have received significant attention for their potential to enhance the capacity and coverage of wireless networks.

Unmanned aerial vehicles (UAVs) enabled wireless networks have been proved as a benefit of mitigating challenges emerging in the conventional terrestrial cellular networks~\cite{liu2019trajectory}. UAVs are capable of being flexibly deployed as aerial BSs in temporary tele-traffic hotspots weaved by political rallies, sporting events or after natural disasters for providing uninterrupted and ubiquitous connectivity~\cite{yan2019comprehensive,mozaffari2019tutorial}. However, before fully reaping the benefits of integrating UAVs into wireless networks, some of the weaknesses of UAVs such as their limited coverage area, meagre energy supply has to be mitigated. By integrating RISs in UAV-enabled wireless networks, concatenated virtual LoS links\footnote[2]{'concatenated link' is constituted by the LoS link between UAVs and MUs and the concatenated link between the RIS and MUs, which is also a LoS link.} between UAVs and mobile users (MUs) can be formed via passively reflecting the incident signals, which lead to extended coverage area as well as less movement of UAVs.

\subsection{State-of-the-art}

\subsubsection{Reinforcement learning in UAV-enabled wireless networks}

It has been proved that integrating UAVs in wireless networks is envisioned to be a promising candidate technique for enhancing the quality of wireless connectivity~\cite{wu2018joint,mozaffari2017mobile}. In an effort to overcome the high dynamic stochastic environment, machine learning (ML) approaches have been invoked in UAV-enabled wireless networks. The authors of~\cite{liu2018energy} leveraged a deep deterministic policy gradient (DDPG) based algorithm for maximizing the defined energy efficiency (EE) function, which considers communications coverage, fairness, energy consumption and connectivity. In~\cite{cui2019multi}, a multi-agent reinforcement learning (RL) algorithm was proposed for tackling the dynamic resource allocation problem in multi-UAV aided wireless networks. The long-term resource allocation problem was formulated as a stochastic game and solved by the proposed RL algorithm. The authors of~\cite{liu2019distributed} presented a decentralized deep reinforcement learning (DRL) based framework for designing the trajectory of a swarm of UAVs. Thus, the geographical fairness of all considered point-of-interests was maximized while the total energy dissipation was minimized. In~\cite{liu2019trajectory}, the trajectory and transmit power of multiple UAVs was designed to maximize the total throughput of all MUs by considering user movement. The formulated problem was solved by a multi-agent RL algorithm. It was demonstrated in~\cite{liu2019trajectory} that the proposed algorithms outperformed the benchmarks when taking into account of user mobility.

\subsubsection{Reinforcement learning in RIS-enhanced wireless networks}

In an effort to effectively exploit RISs for optimizing wireless systems, preliminary researches have tackled some fundamental problems, including channel estimation/modeling, active beamforming for the BS, passive beamforming design for RISs, and resource allocation from the BS to MUs. Efficient optimization approaches, such as convex optimization~\cite{guo2019weighted}, iterative algorithm~\cite{wu2019intelligent}, gradient descent approach~\cite{huang2019reconfigurable}, and alternating algorithm~\cite{shen2019secrecy} have been adopted for tackling the aforementioned challenges. In order to reap the benefits of integrating RISs in the wireless systems, joint active beamforming and passive phase shift design of RIS-enhanced system have been investigated in multiple-input-single-output (MISO) system~\cite{huang2020reconfigurable,feng2020deep}, OFDM-based system~\cite{taha2020deep}, wireless security system~\cite{Helin2020Deep} and Millimeter Wave system~\cite{zhang2020millimeter} with the aid of RL algorithms. In contrast to the alternating optimization techniques, which alternatively optimize the active beamforming at BSs and the passive beamforming at RISs, the RL-based solution is capable of jointly obtaining the beamforming matrix and phase shift matrix. More explicitly, the authors of~\cite{huang2020reconfigurable} invoked a DDPG based algorithm for maximizing the throughput by utilizing the sum rate as the instant rewards for training the DDPG model. In the proposed model, the continuous transmit beamforming and phase shift were jointly optimized with a low complexity. The authors of~\cite{taha2020deep} proposed a DRL based algorithm for maximizing the achievable transmit rate by directly optimizing interaction matrices from the sampled channel knowledge. In the proposed DRL model, only one beam was utilized for each training episode. Thus, the training overhead was avoided, while the dataset collection phase was not required. The authors of~\cite{zhang2020millimeter} proposed a DRL based algorithm for maximizing the throughput with both perfect channel state information (CSI) and imperfect CSI. A quantile regression method was applied for modeling a return distribution for each state-action pair, which modeled the intrinsic randomness in the Markov Decision Process (MDP) interaction between the IR and communication environment. The authors of~\cite{Helin2020Deep} considered the application of RIS in the physical layer security. The system secrecy rate was maximized with the aid of DRL model by designing both active and passive beamforming matrices under users' different quality-of-service (QoS) requirements and time-varying channel condition. Additionally, post-decision state and prioritized experience replay schemes were adopted for enhancing the training performance and secrecy performance.

\subsubsection{NOMA in UAV-enabled/RIS-enhanced wireless networks}

Sparked by the concept of superimposing the signals of multiple associated users at different power levels to exploit the spectrum more efficiently by opportunistically exploring the users' different channel conditions~\cite{liu2017non}, power-domain non-orthogonal multiple access (NOMA) has been invoked for improving the spectrum efficiency and massive connectivity of UAV-enabled/RIS-enhanced wireless networks. In terms of integrating NOMA technique in UAV-enabled wireless networks, the authors~\cite{zhang2020energy} considered the resource allocation problem in NOMA-UAV networks to maximize the EE. An successive convex approximation (SCA) algorithm was leveraged to convert the non-convex challenging problem to a convex problem. In~\cite{lu2020uav}, the UAV-enabled uplink NOMA network was studied for overcoming the inherent latency. The sum rate of all users was maximized by jointly designing the trajectory and power control of the UAV under the QoS constraint. As for integrating NOMA technique in RIS-enhanced wireless systems, the authors of~\cite{liu2020ris} considered the deployment and passive beamforming design problem of the RIS in the MISO-NOMA network. A ${{\text{D}}^{\text{3}}}{\text{QN}}$ algorithm was proposed for tackling the formulated energy efficiency maximizing problem. Simulation results of~\cite{liu2020ris} showed that the EE of the systems can be improved with the aid of the RIS, while NOMA-RIS case outperformed OMA-RIS case in terms of EE. In~\cite{mu2019exploiting}, the downlink NOMA-MISO-RIS network was investigated via jointly designing both the active and passive beamforming. The objective function was formulated as maximizing the sum rate of all MUs. An SCA approach was invoked to design the active and passive beamforming in an alternate manner.

\subsection{Motivations}

Due to the fact that UAVs are battery powered, energy consumption is one of the most important challenges in commercial and civilian applications. The limited endurance of UAVs (usually under 30 minutes) hampers the practical implementation of the UAVs. Before fully reap the advantages of UAV-enabled wireless systems, how to overcome the energy-hungry issue is supposed to be considered. The total energy dissipated by the UAV consists of two aspects, namely, the communication-related and the propulsion-related energy consumption. The first component is dissipated for radiation, signal procession and hardware circuitry, while the other one is dissipated for supporting the hovering and mobility of the UAV. It is worth noting that, the propulsion-related energy consumption accounted for the vast majority of the sum energy consumption (usually more than 95\%), which emphasize the importance of considering the propulsion-related energy consumption model of UAVs when aiming for designing environment friendly wireless networks. In this paper, when we are calculating the energy consumption of the UAV, we only consider the energy dissipated for supporting the hovering and mobility of the UAV.



Since all MUs are roaming continuously, the UAV has to be periodically repositioned according to the mobility of users for establishing LoS wireless links. By deploying an RIS, one can adjust the phase shift matrix of RISs instead of controlling the movement of UAVs for forming concatenated virtual LoS links between the UAV and MUs. Therefore, the UAV can maintain hovering status only when the concatenated virtual LoS links cannot be formed even with the aid of the RIS. By invoking the aforementioned protocol, the total energy dissipation of the UAV is minimized, which in turn, maximizes the endurance of the UAV.

In this paper, both the UAV and MUs are considered as roaming instead of static. Thus, the considered RIS-UAV network is naturally a dynamic scenario, which is challenging for the conventional optimization approaches. Additionally, the goal of deploying and designing the UAV-RIS is for maximizing the long-term benefits instead of current benefits, which falls into the field of the RL algorithm for the reason that this algorithm can incorporate farsighted system evolution instead of myopically optimizing current benefits. Moreover, by integrating NOMA technique in the dynamic RIS-UAV scenario, the decoding order cannot be determined directly by the order of MUs' channel gains. In an effort to guarantee successful successive interference cancelation (SIC), another decoding order constraint based on decoding rate needs to be satisfied while dynamic decoding order needs to be re-determined at each timeslot. Therefore, joint trajectory design of the UAV, passive beamforming design at the RIS, decoding order and power allocation determination are expected to be optimized. The search-space is expanded as the number of parameters increases, which makes the conventional gradient-based optimization techniques unsuitable. Against the aforementioned background, RL algorithm, which is a powerful AI paradigm that is capable of empowering agents by learning from the dynamic environment, was selected as the methodology for tackling problems in RIS-UAV networks. Since RISs enjoy discrete phase shifts, the state space and action space are also discrete. Thus, the policy-based algorithm, which aims for solving problems with continuous state space and action space, is not suitable for controlling RISs. the deep Q-network (DQN) algorithm can be used for supporting the UAV/RIS (agents) in their interactions with the environment (states), whilst finding the optimal behavior (actions) of the UAV/RIS. Thus, it is applied for solving challenging problems in the UAV-RIS enhanced wireless networks.

\subsection{Contributions}

The contributions of this paper are as follows:

\begin{itemize}
\item We propose a novel UAV-RIS framework for attaining the long-term benefits of the network controlling the RIS and designing the trajectory of the UAV in RIS-enhanced UAV-enabled wireless networks, where an RIS is deployed for enhancing the wireless connectivity and reduce the movement of the UAV. Additionally, we formulate the energy consumption minimization problem by jointly deciding the movement of the UAV and the passive beamforming at the RIS.
\item We investigate both OMA case and NOMA case in RIS-enhanced UAV-enabled wireless networks, zero-forcing based linear precoding method is employed for eliminating the multiuser interference in OMA case and the inter-cluster interference in NOMA case, respectively. Additionally, in contrast to the conventional MISO-NOMA networks, we consider dynamic decoding order in NOMA-RIS-UAV case due to the mobility of both UAV and MUs.
\item We adopt a decaying learning rate deep Q-network (D-DQN) based algorithm to tackle the formulated joint trajectory and phase shift design problem. In contrast to the conventional DQN algorithm, decaying learning rate is leveraged in the proposed D-DQN based algorithm for attaining a tradeoff between accelerating training speed and converging to the local optimal, as well as for avoiding oscillation. The central controller is selected as an agent for determining the movement of the UAV, as well as the passive beamforming design. Additionally, we adopt the three-dimensional (3D) radio map for verifying the performance of the proposed D-DQN based algorithm.
\item We demonstrate that the proposed D-DQN based algorithm can converge under minor constraint. With the aid of the RIS, the energy dissipation of the UAV is capable of being reduced by roughly 23.3\%, while RIS-NOMA case consumes 11.7\% less energy than RIS-OMA case.
\end{itemize}

\subsection{Organization and Notations}

The remainder of this paper is structured as follows. Section II focuses on the system model and problem formulation of energy consumption minimization for the UAV. Section III elaborates on the proposed D-DQN based algorithms for solving the formulated problem. The simulation results are illustrated in Section V. Finally, Section VI concludes the main concept, insights and results of this paper. The list of notations is illustrated in Table~\ref{List of Notations}.

\begin{table}[htbp]\footnotesize
\caption{List of Notations}
\centering
\begin{tabular}{|l||r|l||r|}
\hline
\centering
 Notations & Description & Notations & Description\\
\hline
\centering
 $M$ & Antenna number of the UAV & $N$ & Reflecting elements of the RIS\\
\hline
\centering
 $K$ & Number of users & $L$ & Number of clusters\\
\hline
\centering
 $P_l$ & Transmit power for the $l$-th cluster & ${\boldsymbol{H}}_{B,S} \in {\mathbb{C}^{N \times M}}$ & Channel metric of the UAV-RIS link\\
\hline
\centering
$\boldsymbol{h}_{S,k}^{H} \in {\mathbb{C}^{1 \times N}}$ & Channel metric of the RIS-MU link &  $\boldsymbol{h}_{B,k}^{H} \in {\mathbb{C}^{1 \times M}}$ & Channel metric of the UAV-MU link\\
\hline
\centering
 ${x_{l}}$ & Transmit signal of the $l$-th cluster & $y_{l,i}$ & Received signal of user $i$ in the $l$-th cluster\\
\hline
\centering
 $N_0$ & Noise power spectral & $B$ & Bandwidth \\
\hline
\centering
 ${\theta _{S,n}}$ & Phase shift & $\beta _{S,n}$ & Amplitude reflection coefficient\\
\hline
\centering
 $\alpha $ & Path loss exponent & ${R_{l,i}}$ & Achievable rate of user $i$ in the $l$-th cluster\\
\hline
\centering
 $N_x$ & Size of neuron reservoir & $s_t$ & State in D-DQN algorithm\\
 \hline
\centering
 $a_t$ & Action in D-DQN algorithm & $r_t$ & Reward in D-DQN algorithm\\
 \hline
\end{tabular}
\label{List of Notations}
\end{table}

\section{System Model and Problem Formulation}

\subsection{System Model}

We consider the downlink MISO communications in a particular area, where terrestrial infrastructures are destroyed due to natural disaster or had not been installed. As illustrated in Fig.1, an UAV equipped with $M$ antenna is employed for providing wireless service for a total number of $K$ single-antenna users. In this paper, all MUs are considered as roaming around in this area. In an effort to enhance the quality of wireless service by forming concatenated virtual LoS propagation between the UAV and the users via passively reconfiguring their incident signals, it is assumed that an RIS is employed on the facade of a particular high-rise building~\cite{di2019smart,qingqing2019towards,wu2018intelligent}, while the RIS is equipped with a number of $N$ reflecting elements.

\begin{figure} [t!]
\centering
\includegraphics[width=4.5in]{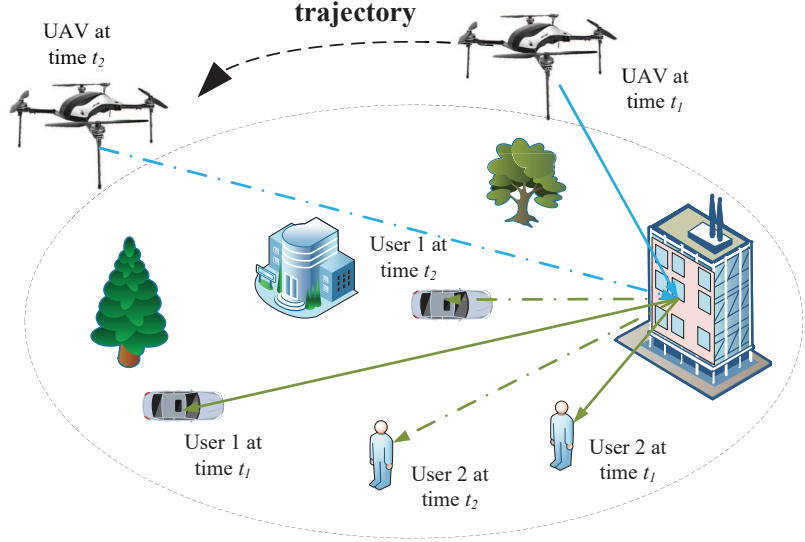}
\caption{Illustration of RIS in UAV-enabled wireless networks.}\label{scenario}
\end{figure}

\begin{remark}\label{remark:motivation}
Since all users are roaming continuously, the UAV has to be periodically repositioned according to the mobility of users for establishing LoS wireless links. With the aid of RIS, one can adjust the phase shift of the RIS instead of controlling the movement of UAV for forming concatenated virtual LoS propagation between the UAV and the users, which, in turn, leads to the reduction of UAV's energy dissipation.
\end{remark}

In contrast to the conventional MISO networks, the received signal consists of two parts, namely signals of direct link (UAV-MU link) and reflecting link (UAV-RIS-MU link). Denote ${\boldsymbol{H}}_{U,S} \in {\mathbb{C}^{N \times M}}$, $\boldsymbol{h}_{S,k}^{H} \in {\mathbb{C}^{1 \times N}}$ as the channel of UAV-RIS link, $\boldsymbol{h}_{S,k}^{H} \in {\mathbb{C}^{1 \times N}}$ as the channel of RIS-MU link, and $\boldsymbol{h}_{U,k}^{H} \in {\mathbb{C}^{1 \times M}}$ as the channel of UAV-MU link, while $k \in \mathcal{K},{\kern 1pt} {\kern 1pt} {\kern 1pt} \left| \mathcal{K} \right| = K$ and $n \in \mathcal{N},{\kern 1pt} {\kern 1pt} {\kern 1pt} \left| \mathcal{N} \right| = N$. Denote ${{\boldsymbol{h}}^H}$ as the conjugate transpose of matrix $\boldsymbol{h}$.

Denote ${{\boldsymbol{\Theta }}} = [{\Theta _1}, \cdots ,{\Theta _n}, \cdots ,{\Theta _N}]$ and ${\Theta _n} = {\beta _n}{e^{j{\theta _n}}}$, where ${\theta _{n}} \in \left[ {0,2\pi } \right]$ represents the phase shift, while ${\beta _{n}} \in \left[ {0,1} \right]$ denotes the amplitude reflection coefficient~\cite{guo2019weighted,abeywickrama2019intelligent}. Before design the passive phase shift of the RIS, the active beamforming design for the UAV has to be considered. In this paper, we adopt a zero-forcing precoding method for designing the active beamforming at the UAV. Our future work will consider the problem of jointly optimizing the active and passive beamforming.

\subsection{Channel Model}

In terms of the channel model between the UAV and MUs, we consider the UAV channel model provided by the 3GPP specifications~\cite{muruganathan2018overview}, in which the path loss is stochastically determined by LoS and non-line-of-sight (NLoS) link states. The LoS and NLoS propagation are decided by the UAV's altitude, the distance between the UAV and the associated users, as well as the carrier frequency. The path loss for the UAV-MU link of MU $k$ can be expressed as
\begin{align}\label{lk}
{{L_k}= \left\{ {\begin{array}{*{20}{c}}
  {30.9 + \left( {22.25 - 0.5{{\log }_{10}}{h}} \right){{\log }_{10}}d_k + 20{{\log }_{10}}{f_c},{\kern 1pt} {\kern 1pt} {\kern 1pt} {\kern 1pt} {\kern 1pt} {\text{if}}{\kern 1pt} {\kern 1pt} {\text{LoS}}{\kern 1pt} {\kern 1pt} {\kern 1pt} {\kern 1pt} {\text{link}},} \\
  {\max \left\{ {L_k^{{\text{LoS}}},32.4 + \left( {43.2 - 7.6{{\log }_{10}}{h}} \right){{\log }_{10}}d_k + 20{{\log }_{10}}f_c} \right\},{\kern 1pt} {\kern 1pt} {\kern 1pt} {\kern 1pt} {\kern 1pt} {\text{if}}{\kern 1pt} {\kern 1pt} {\text{NLoS}}{\kern 1pt} {\kern 1pt} {\kern 1pt} {\kern 1pt} {\text{link}},}
\end{array}} \right.}
\end{align}
where $h$ denotes the altitude of the UAV, $f_c$ represents the carrier frequency, ${L_k^{{\text{LoS}}}}$ is the path loss for the LoS link.

The LoS propagation state are stochastically determined by the LoS probability ${P_{{\text{LoS}}}}$, described as
\begin{align}\label{plos}
{{P_{{\text{LoS}}}} = \left\{ {\begin{array}{*{20}{c}}
  {1,}&{{\text{if}}{\kern 1pt} {\kern 1pt} {\kern 1pt} {\kern 1pt} {\kern 1pt} \sqrt {d_k^2 - {h^2}}  \le {d_0}}, \\
  {\frac{{{d_0}}}{{\sqrt {d_k^2 - {h^2}} }} + \exp \left\{ {\left( {\frac{{ - \sqrt {d_k^2 - {h^2}} }}{{{p_1}}}} \right)\left( {1 - \frac{{{d_0}}}{{\sqrt {d_k^2 - {h^2}} }}} \right)} \right\},}&{{\text{if}}{\kern 1pt} {\kern 1pt} {\kern 1pt} {\kern 1pt} {\kern 1pt} \sqrt {d_k^2 - {h^2}}  > {d_0}},
\end{array}} \right.}
\end{align}
where ${d_0} = \max [294.05{\log _{10}}{h} - 432.94,18]$, while ${p_1} = 233.98{\log _{10}}{h} - 0.95$. Thus, the NLoS probability can be calculated as ${P_{{\text{NLoS}}}} = 1 - {P_{{\text{LoS}}}}$.

In equation \eqref{lk}, the distance from the UAV to MU $k$ at time $t$ can be calculated as
\begin{align}\label{dt}
{{d_k} = \sqrt {{h^2} + {{\left[ {x_{\text{UAV}} - {x_k}} \right]}^2} + {{\left[ {y_{\text{UAV}} - {y_k}} \right]}^2}} }.
\end{align}

The channel between the UAV and MU $k$ at time $t$ is given by
\begin{align}\label{gt}
{G_k = {h_k} \cdot {10^{{{ - {L_k}} \mathord{\left/
 {\vphantom {{ - {L_k}} {10}}} \right.
 \kern-\nulldelimiterspace} {10}}}}},
\end{align}
where ${h_k}$ represents the fading coefficient.

It is worth noting that the acquisition of timely and accurate CSI becomes challenging due to the reason that RISs are not capable of performing active transmission/reception and signal processing. Hence our future work would consider investigating the open problem of CSI acquisition. The UAV-RIS channel is similar generated by following the above procedure of the UAV-MU channel. In terms of the channel model of RIS-MU link, the path loss from the RIS to MUs can be modeled as $\eta \left( d_{\text{RIS-MU}} \right) = {C_0}{\left( {\frac{d_{\text{RIS-MU}}}{{{d_0}}}} \right)^{ - \alpha_{\text{RIS-MU}} }}$~\cite{wu2018intelligent}, where $C_0$ represents the path loss in the condition of $d_0=1$ meters (m), $d_{\text{RIS-MU}}$ denotes the distance between the RIS and MUs, and $\alpha_{\text{RIS-MU}}$ represents the path loss exponent.

\subsection{3D radio map reconstruction and user mobility}

In this subsection, we model the urban city environment as a set of buildings, where each building is modeled as a set of cubes. As shown in Fig.2, a cylindrical building can be geometrically constructed by a large number of long cubes, while each cube can be described by 4 line segments and its height. In this case, we can determine whether the UAV-MU link is blocked (NLoS) by calculating if the line segment joining the UAV and the MU penetrate one of these cubes.

\begin{remark}\label{remark:radiomap}
Given the coordinates of the UAV and a particular MU, the link between them can be distinguished as LoS or NLoS propagation.
\end{remark}

\begin{figure}[http]
	\begin{minipage}[t]{0.45\textwidth}
		\centering
		\includegraphics[scale=0.5]{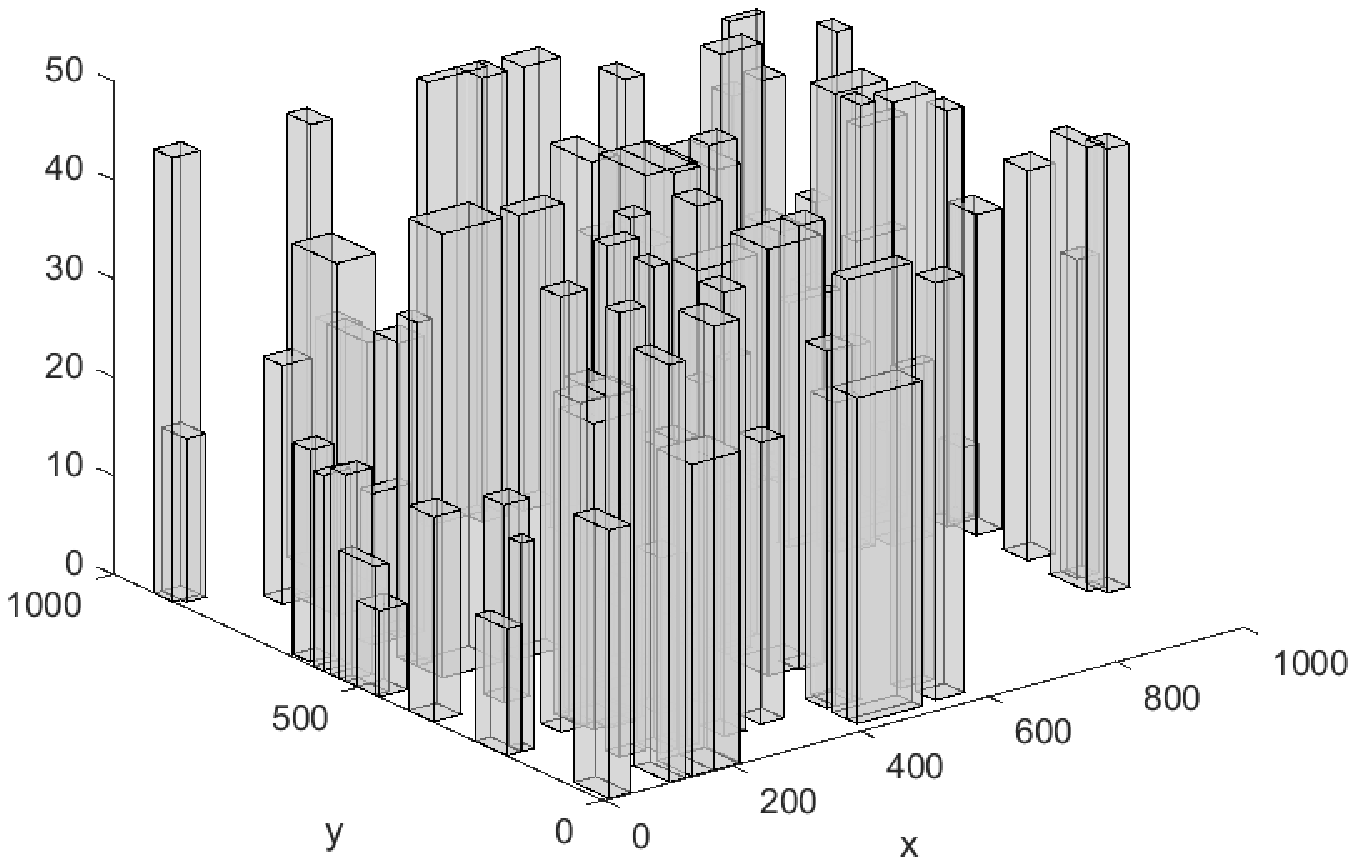}
		\caption{3D radio map of a dense urban area.\label{citymap}}
	\end{minipage}
	\qquad
	\begin{minipage}[t]{0.45\textwidth}
		\centering
		\includegraphics[scale=0.5]{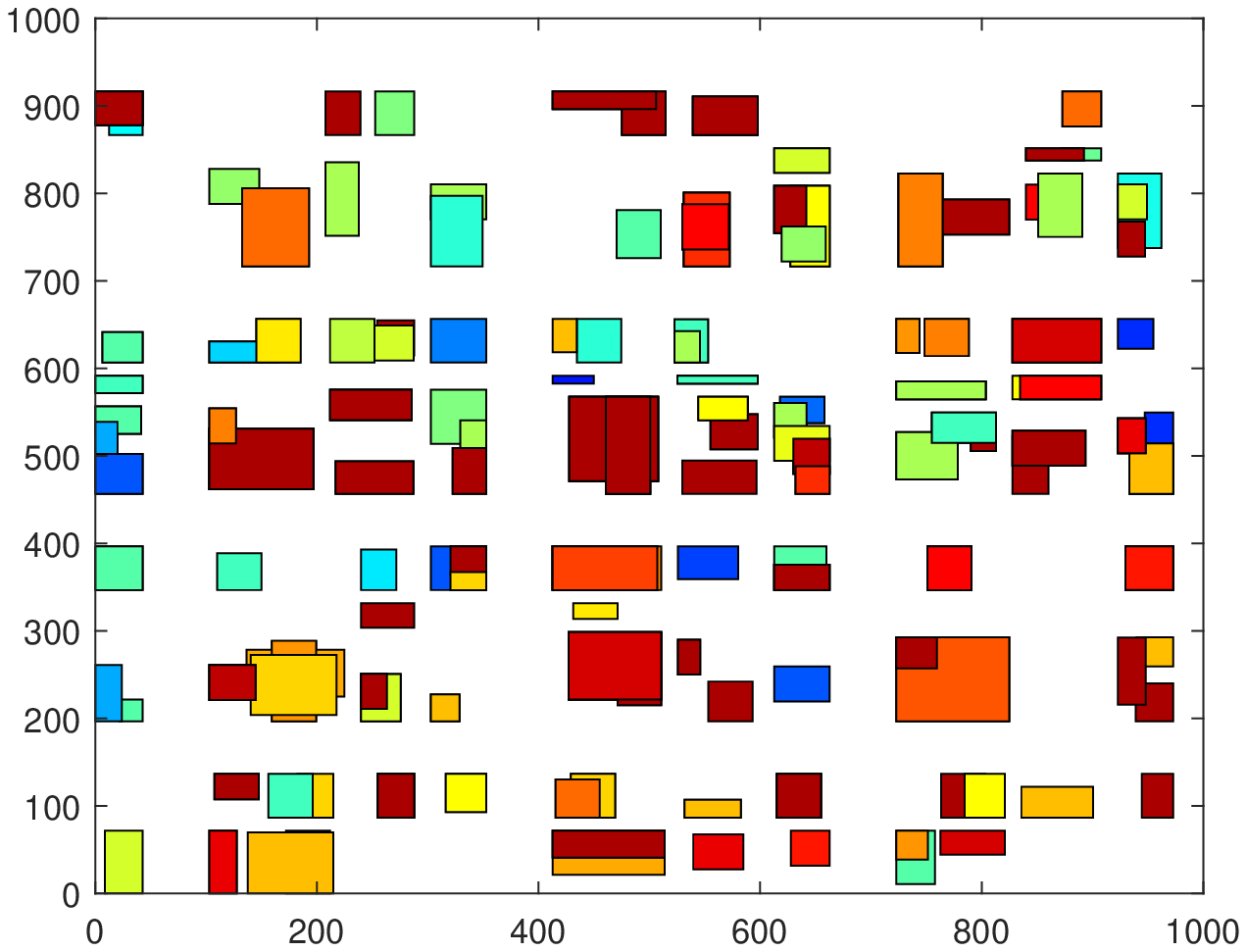}
		\caption{2D radio map of a dense urban area.\label{citymap2D}}
	\end{minipage}
\end{figure}

In terms of the user mobility, the wide use of online social networks over smart-phones has accumulated a rich set of geographical data that describes the anonymous users' mobility in the physical world. The users' locations can be predicted by mining data from social networks, given that the observed movement is associated with certain reference locations. One of the most effective method of collecting position information relies on the Twitter API. When Twitter users tweet, their GPS-related position information is recorded by the Twitter API and it becomes available to the general public~\cite{liu2019trajectory}. Indeed, the prediction accuracy of the users' location is essential for the design of RIS-UAV enabled wireless networks. Collecting and processing data gleaned from online social networks is of lower complexity than other localization methods, such as channel estimation. Thus, we can strike an accuracy vs complexity tradeoff. Additionally, the prediction accuracy can be improved by collecting a user's current position more frequently, which requires for additional communication resources.

\subsection{Energy Dissipation Model for the UAV}

The total energy dissipated by a rotary-wing UAV consists of two aspects, the communication-related energy consumption and the propulsion-related energy consumption. The first component is dissipated for radiation, signal procession and hardware circuitry, while the other one is dissipated for supporting the hovering and mobility of the UAV. It is worth noting that, the propulsion-related energy consumption accounted for the vast majority of the total energy consumption(Usually more than 95\%). For the rotary-wing UAV, the energy consumption is related to its velocity and acceleration.

In terms of the energy dissipation model for a rotary-wing UAV, the propulsion energy dissipation is given by~\cite{zeng2019energy}
\begin{align}\label{rotary1}
{\begin{gathered}
  \bar E(t) = \frac{1}{{{T_r}}} \cdot \sum\limits_{t = n{T_r}}^{(n + 1){T_r}} {{E_b}\left( {1 + \frac{{3{v^2}(t)}}{{U_{{\text{tip}}}^2}}} \right)}  + \frac{1}{{{T_r}}} \cdot \sum\limits_{t = n{T_r}}^{(n + 1){T_r}} {{E_i}\left( {\sqrt {1 + \frac{{{v^4}(t)}}{{4v_0^4}}}  - \frac{{{v^2}(t)}}{{2v_0^2}}} \right)}  \hfill \\
  {\kern 1pt} {\kern 1pt} {\kern 1pt} {\kern 1pt} {\kern 1pt} {\kern 1pt} {\kern 1pt} {\kern 1pt} {\kern 1pt} {\kern 1pt} {\kern 1pt} {\kern 1pt} {\kern 1pt} {\kern 1pt} {\kern 1pt} {\kern 1pt} {\kern 1pt} {\kern 1pt} {\kern 1pt} {\kern 1pt} {\kern 1pt} {\kern 1pt} {\kern 1pt} {\kern 1pt} {\kern 1pt} {\kern 1pt} {\kern 1pt}  + \frac{1}{{{T_r}}} \cdot \sum\limits_{t = n{T_r}}^{(n + 1){T_r}} {\frac{1}{2}{f_d}\rho sA} {v^3}(t), \hfill \\
\end{gathered} }
\end{align}
where ${E_b} = \frac{\delta }{8}\rho sA{\Omega ^3}{R^3}$ represents the blade profile power of the UAV in hovering status with $s$ denoting the fuselage drag ratio and rotor solidity, $\rho $ representing the air density, $\delta$ is the profile drag coefficient, $A$ implying the total propeller area, while $\Omega $ indicates the blade angular velocity, $R$ is the full rotor blade of radius. $E_i=(1 + \kappa )\frac{{{W^{{3 \mathord{\left/
 {\vphantom {3 2}} \right.
 \kern-\nulldelimiterspace} 2}}}}}{{\sqrt {2\rho A} }}$ denotes the induced power of the UAV in hovering status with $\kappa $ denoting the incremental correction factor to induced power, $W$ denoting the weight of the UAV. Finally, $v_0$ is mean rotor induced velocity in hovering status, $f_d$ denotes the fuselage drag ratio.

\subsection{Signal Model for OMA Scheme}

By considering the linear transmit pre-coding at the UAV, the transmitting signal from the UAV in OMA case are given by
\begin{align}\label{transmitsignalOMA}
{{\boldsymbol{x}} = \sum\limits_{k = 1}^K {\sqrt {{p_k}} {\boldsymbol{g}_k}{s_k}} },
\end{align}
where $s_k$ represents the transmitted data symbol for the $k$-th MU with with $k \in \mathcal{K},{\kern 1pt} {\kern 1pt} {\kern 1pt} \left| \mathcal{K} \right| = K$, ${\boldsymbol{g}_k} \in {\mathbb{C}^{M \times 1}}$ denotes the corresponding transmit beamforming vector, and $p_k$ represents the allocated transmit power from the UAV to the $k$-th MU.

The received signal of MU $k$ is given by
\begin{align}\label{receivesignalOMA}
{{y_k} = \left( {{\boldsymbol{h}}_{U,k}^{H} + {\boldsymbol{h}}_{S,k}^{H}{{\boldsymbol{\Theta }}}{\boldsymbol{H}}_{U,S}} \right)\sum\limits_{j = 1}^K {\sqrt {{p_j}} {{\boldsymbol{g}}_j}{s_j}}  + {n_k}},
\end{align}
where ${n_k} \sim \mathcal{C}\mathcal{N}\left( {0,\sigma _k^2} \right)$ represents the additive white Gaussian noise (AWGN). Based on \eqref{receivesignalOMA}, the received signal-to-interference-plus-noise (SINR) of MU $k$ can be calculated as follow
\begin{align}\label{SINROMA}
{{\gamma _k} = \frac{{{p_k}{{\left| {\left[ {{\boldsymbol{h}}_{U,k}^{H} + {\boldsymbol{h}}_{S,k}^{H}{{\boldsymbol{\Theta }}}{\boldsymbol{H}}_{U,S}} \right]{{\boldsymbol{g}}_k}} \right|}^2}}}{{\sum\limits_{j \ne k}^K {{p_j}{{\left| {\left[ {{\boldsymbol{h}}_{U,j}^{H} + {\boldsymbol{h}}_{S,j}^{H}{{\boldsymbol{\Theta }}}{\boldsymbol{H}}_{U,S}} \right]{{\boldsymbol{g}}_j}} \right|}^2} + \sigma _k^2} }}}.
\end{align}

To eliminate the multiuser interference, zero-forcing (ZF)-based linear pre-coding method is invoked at the UAV~\cite{huang2019reconfigurable}\footnote[3]{In this paper, the active beamforming is decided by ZF precoding method. Optimizing the precoding matrix by machine learning algorithms will be considered in our future work.}. Denote ${\boldsymbol{h}}_j^H,{\kern 1pt} {\kern 1pt} {\kern 1pt} j \in \mathcal{K}$ be the combined channel of the $j$-th MU, the precoding metrics can be expressed as
\begin{align}\label{ZFOMA1}
{\left\{ {\begin{array}{*{20}{c}}
  {\left[ {{\boldsymbol{h}}_{U,j}^{H} + {\boldsymbol{h}}_{S,j}^{H}{{\boldsymbol{\Theta }}}{\boldsymbol{H}}_{U,S}} \right]{{\boldsymbol{g}}_k} = 0,}&{{\kern 1pt} \forall j \ne k,{\kern 1pt} {\kern 1pt} {\kern 1pt} j \in \mathcal{K},} \\
  {\left[ {{\boldsymbol{h}}_{U,k}^{H}(t) + {\boldsymbol{h}}_{S,k}^{H}{{\boldsymbol{\Theta }}}{\boldsymbol{H}}_{U,S}} \right]{{\boldsymbol{g}}_k} = 1,}&{j = k.}
\end{array}} \right.}
\end{align}

Denote ${{\boldsymbol{H}}^H} = {\boldsymbol{H}}_{\text{UAV-MU}}^H + {\boldsymbol{H}}_{\text{RIS-MU}}^H{{\boldsymbol{\Theta }}_l}{\boldsymbol{H}}_{\text{UAV-RIS}}^H$, where we have ${\boldsymbol{H}}_{\text{UAV-MU}}^H = {\left[ {{\boldsymbol{h}}_{U,1}, \cdots ,{\boldsymbol{h}}_{U,K}} \right]^H}$ and ${\boldsymbol{H}}_{\text{RIS-MU}}^H = {\left[ {{\boldsymbol{h}}_{S,1}, \cdots ,{\boldsymbol{h}}_{S,K}} \right]^H}$. Thus, the transmit pre-coding metric ${\boldsymbol{G}} = \left[ {{{\boldsymbol{g}}_1}, \cdots ,{{\boldsymbol{g}}_K}} \right]$ can be derived by the pseudo-inverse of the combined channel ${{\boldsymbol{H}}^H}$, which can be expressed as
\begin{align}\label{ZFOMA2}
{{\boldsymbol{G}} = {\boldsymbol{H}}{\left( {{{\boldsymbol{H}}^H}{\boldsymbol{H}}} \right)^{ - 1}}}.
\end{align}

Thus, the instantaneous transmit rate of the $k$-th MU at timeslot $t$ in OMA case can be expressed as
\begin{align}\label{transmitrateOMA}
{{R_k} = {B_k}{\log _2}\left( {1 + {\gamma _k}} \right) = {B_k}{\log _2}\left( {1 + \frac{{{p_k}}}{{{\sigma ^2}}}} \right)},
\end{align}
where $B_k$ represents the bandwidth allocated by UAV to the $k$-th MU.

\subsection{Signal Model for NOMA Scheme}

By invoking NOMA technique instead of OMA technique, the spectrum efficiency can be further improved. The transmit signal from the UAV to the $l$-th cluster is given by

\begin{align}\label{transsignalNOMA}
{{x_l}(t) = \sqrt {{\alpha _{l,a}}(t)} {s_{l,a}}(t) + \sqrt {{\alpha_{l,b}}(t)} {s_{l,b}}(t)},
\end{align}
where ${s_{l,a}}$ and ${s_{l,b}}$ represent the signals for MU $a$ and MU $b$ in the same cluster, respectively. Since the concept of NOMA technique is superimposing the signals of two associated users at different power levels, we denote ${\alpha_{l,a}}$ and ${\alpha_{l,b}}$ as the power allocation factors for MU $a$ and MU $b$, respectively. Naturally, the power allocation factors satisfy $\alpha_{l,a}+\alpha_{l,b}=1$. Hence, the received signal of a particular MU $i$ in cluster $l$ can be calculated as
\begin{align}\label{transmitsignalNOMA}
{{y_{l,i}} = \left( {{\boldsymbol{h}}_{U,l,i}^H + {\boldsymbol{h}}_{S,l,i}^H{{\boldsymbol{\Theta }}_S}{{\boldsymbol{H}}_{U,S}}} \right)\sum\limits_{l = 1}^L {{{\boldsymbol{w}}_l}{x_l}}  + {n_{l,i}} },
\end{align}
where ${{{\boldsymbol{w}}_l}}$ represents the precoding vector of cluster $l$.

In the NOMA case, each MU in the same cluster adopts SIC method for removing the intra-cluster interference~\cite{liu2016cooperative,ding2017survey}. The strong MU in the cluster is capable of removing the intra-cluster interference caused by the weak MU with the aid of SIC method. On the other hand, the weak MU is designed to decode the received signal directly without invoking SIC~\cite{liu2018multiple,liu2017non}. Denote MU $a$ as the strong user in cluster $l$. Thus, the received signal of MU $a$ can be expressed as
\begin{align}\label{SINRNOMA}
{\begin{gathered}
  {y_{l,a}} = \left( {{\boldsymbol{h}}_{U,l,a}^H + {\boldsymbol{h}}_{S,l,a}^H{{\boldsymbol{\Theta }}_S}{{\boldsymbol{H}}_{U,S}}} \right){{\boldsymbol{w}}_l}\left( {\sqrt {{\alpha _{l,a}}} {s_{l,a}} + \sqrt {{\alpha_{l,b}}} {s_{l,b}}} \right) \hfill \\
  {\kern 1pt} {\kern 1pt} {\kern 1pt} {\kern 1pt} {\kern 1pt} {\kern 1pt} {\kern 1pt} {\kern 1pt} {\kern 1pt} {\kern 1pt} {\kern 1pt} {\kern 1pt} {\kern 1pt} {\kern 1pt} {\kern 1pt} {\kern 1pt} {\kern 1pt} {\kern 1pt} {\kern 1pt} {\kern 1pt} {\kern 1pt} {\kern 1pt} {\kern 1pt} {\kern 1pt} {\kern 1pt} {\kern 1pt} {\kern 1pt} {\kern 1pt} {\kern 1pt}  + \left( {{\boldsymbol{h}}_{U,l,a}^H(t) + {\boldsymbol{h}}_{S,l,a}^H{{\boldsymbol{\Theta }}_S}{{\boldsymbol{H}}_{U,S}}} \right)\sum\limits_{j = 1,j \ne l}^L {{{\boldsymbol{w}}_j}{x_j}}  + {n_{l,a}}, \hfill \\
\end{gathered} }
\end{align}
where $\left( {{\boldsymbol{h}}_{U,l,a}^H + {\boldsymbol{h}}_{U,l,a}^H{{\boldsymbol{\Theta }}_S}{{\boldsymbol{H}}_{U,S}}} \right)\sum\limits_{j = 1,j \ne l}^L {{{\boldsymbol{w}}_j}{x_j}} $ denotes the inter-cluster interference in multi-cell NOMA networks, $\left( {{\boldsymbol{h}}_{U,l,a}^H + {\boldsymbol{h}}_{S,l,a}^H{{\boldsymbol{\Theta }}_S}{{\boldsymbol{H}}_{U,S}}} \right){{\boldsymbol{w}}_l}\sqrt {{a_{l,b}}} {s_{l,b}}$ represents the intra-cluster interference.

In the same manner with OMA case, ZF precoding approach is leveraged for eliminating the inter-cluster interference for the strong user~\cite{huang2019reconfigurable,basar2019large,huang2018energy}. Although the dirty paper coding (DPC) is proved to achieve the maximum capacity in multi-user MIMO-NOMA system, it is non-trivial to be implemented in practice for the reason that it adopts brute-force searching. Thus, ZF-based linear precoding method, which is of a low complexity is employed. The corresponding ZF pre-coding constraints can be expressed as
\begin{align}\label{ZFNOMA1}
{\left\{ {\begin{array}{*{20}{c}}
  {\left[ {{\boldsymbol{h}}_{U,j}^{H} + {\boldsymbol{h}}_{S,j}^{H}{{\boldsymbol{\Theta }}_S}{\boldsymbol{H}}_{U,S}} \right]{{\boldsymbol{w}}_l} = 0,}&{{\kern 1pt} \forall j \ne l,{\kern 1pt} {\kern 1pt} {\kern 1pt} j \in \mathcal{L},} \\
  {\left[ {{\boldsymbol{h}}_{U,l}^{H} + {\boldsymbol{h}}_{S,l}^{H}{{\boldsymbol{\Theta }}_S}{\boldsymbol{H}}_{U,S}} \right]{{\boldsymbol{w}}_l} = 1,}&{j = l.}
\end{array}} \right.}
\end{align}

The same like OMA case, the optimal transmit pre-coding metric ${\boldsymbol{W}} = \left[ {{{\boldsymbol{w}}_1}, \cdots ,{{\boldsymbol{w}}_L}} \right]$ can also be calculated by the pseudo-inverse of the combined channel ${{\boldsymbol{H}}^H}$, which can be expressed as
\begin{align}\label{ZFNOMA2}
{{\boldsymbol{W}} = {\boldsymbol{H}}{\left( {{{\boldsymbol{H}}^H}{\boldsymbol{H}}} \right)^{ - 1}}}.
\end{align}

With the aid of ZF precoding method, the inter-cluster interference suffered by the strong users can be removed, while the intra-user interference can also be removed with the aid of successful SIC. However, the weak still suffers the inter-cluster interference. Therefore, the received signal of the strong user can be calculated as
\begin{align}\label{yl1NOMA}
{{y_{l,a}}{\kern 1pt}  = \left( {{\boldsymbol{h}}_{U,l,a}^H + {\boldsymbol{h}}_{S,l,b}^H{{\boldsymbol{\Theta }}_S}{{\boldsymbol{H}}_{U,S}}} \right){{\boldsymbol{w}}_l}\sqrt {{\alpha_{l,a}}} {s_{l,a}} + {n_{l,a}}},
\end{align}
while the received signal of the weak user can be calculated as
\begin{align}\label{yl2NOMA}
{\begin{gathered}
  {y_{l,b}}{\kern 1pt}  = \left( {{\boldsymbol{h}}_{U,l,b}^H + {\boldsymbol{h}}_{S,l,b}^H{{\boldsymbol{\Theta }}_S}{{\boldsymbol{H}}_{U,S}}} \right){{\boldsymbol{w}}_l}\left( {\sqrt {{\alpha_{l,a}}} {s_{l,a}} + \sqrt {{\alpha_{l,b}}} {s_{l,b}}} \right) \hfill \\
  {\kern 1pt} {\kern 1pt} {\kern 1pt} {\kern 1pt} {\kern 1pt} {\kern 1pt} {\kern 1pt} {\kern 1pt} {\kern 1pt} {\kern 1pt} {\kern 1pt} {\kern 1pt} {\kern 1pt} {\kern 1pt} {\kern 1pt} {\kern 1pt} {\kern 1pt} {\kern 1pt} {\kern 1pt} {\kern 1pt} {\kern 1pt} {\kern 1pt} {\kern 1pt} {\kern 1pt} {\kern 1pt} {\kern 1pt} {\kern 1pt} {\kern 1pt} {\kern 1pt}  + \left( {{\boldsymbol{h}}_{U,l,b}^H + {\boldsymbol{h}}_{S,l,b}^H{{\boldsymbol{\Theta }}_S}{{\boldsymbol{H}}_{U,S}}} \right)\sum\limits_{j = 1,j \ne l}^L {{{\boldsymbol{w}}_j}{x_j}}  + {n_{l,b}}. \hfill \\
\end{gathered} }
\end{align}

Therefore, the received SINR for both strong user and weak user can be expressed as

\begin{align}\label{SINR1NOMA}
{{\gamma _{l,a}} = \frac{{{{\left| {\left( {{\boldsymbol{h}}_{U,l,a}^H + {\boldsymbol{h}}_{S,l,a}^H{{\boldsymbol{\Theta }}_S}{{\boldsymbol{H}}_{U,S}}} \right){{\boldsymbol{w}}_l}\sqrt {{\alpha_{l,a}}} {s_{l,a}}} \right|}^2}}}{{\sigma _l^2}} = \frac{{{\alpha_{l,a}}{P_{l}}}}{{\sigma _l^2}} },
\end{align}
and
\begin{align}\label{SINR2NOMA}
{{\gamma _{l,b}} = \frac{{{{\left| {{{\boldsymbol{h}}_{l,b}}{{\boldsymbol{w}}_l}} \right|}^2}{\alpha_{l,b}}{P_{l}}}}{{{{\left| {{{\boldsymbol{h}}_{l,b}}{{\boldsymbol{w}}_l}} \right|}^2}{\alpha_{l,b}}{P_{l}} + {{\left| {{{\boldsymbol{h}}_{l,b}}\sum\limits_{j = 1,j \ne l}^L {{{\boldsymbol{w}}_l}{x_j}} } \right|}^2} + \sigma _l^2}} },
\end{align}
where ${{\boldsymbol{h}}_{l,b}} = {\boldsymbol{h}}_{U,l,b}^H+ {\boldsymbol{h}}_{S,l,b}^H{{\boldsymbol{\Theta }}_S}{{\boldsymbol{H}}_{U,S}}$.

\begin{remark}\label{remark:decodingorder}
In contrast to the conventional NOMA systems, for MISO-NOMA networks, the decoding order cannot be decided directly by the order of MUs' channel gains, another decoding rate constraint needs to be satisfied for guaranteeing successful SIC. Additionally, by integrating RISs in the MISO-NOMA system, the channel response is also modified by RISs. Thus, the decoding order constraint in this
paper has to be satisfied at each timeslot.
\end{remark}

The decoding rate condition is given by ${\gamma}{_{l,b \to l,a}} \ge {\gamma}_{l,b \to l,b}$, ${\pi _l}(a) \ge {\pi _l}(b)$~\cite{cui2017optimal}, where $\gamma_{l,b \to l,a}$ denotes the SINR of user $a$ to decode user $b$.

Since both the UAV and MUs are considered as roaming continuously, the decoding order has to be re-determined at each timeslot to adapt to the dynamic environment.

\subsection{Problem Formulation}
We are interested in minimizing the energy consumption of the UAV while guaranteeing the wireless service quality from the UAV to the users at each timeslot by jointly optimizing the phase shift metric of the RIS, the movement of the UAV, the power allocation from the UAV to MUs, and the dynamic decoding order. Denote ${{\boldsymbol{\theta}}} = [{\theta _{1}}(t), \cdots ,{\theta _{n}}(t), \cdots ,{\theta _{N}}(t)]$, ${{\boldsymbol{P}}} = [{p _{1}}(t), \cdots ,{p _{k}}(t), \cdots ,{p _{K}}(t)]$ and $\boldsymbol{Q}= {[x_{\text{UAV}}(t),y_{\text{UAV}}(t)]^T}$. Thus, the optimization problem can be formulated as

\begin{center}
\begin{subequations}\label{optimizationproblem2}
\begin{align}
\mathop  {\mathop {\min }\limits_{\boldsymbol{\theta}, \boldsymbol{P}, \boldsymbol{Q}} {\kern 1pt} {\kern 1pt} {\kern 1pt} {E_{{\text{UAV}}}}{\text{ = }}\sum\limits_{t = 0}^T {\bar E(t)} } \\
{\text{s}}{\text{.t}}{\text{.}}{\kern 1pt} {\kern 1pt} {\kern 1pt} {\kern 1pt} {\kern 1pt} {\kern 1pt} {\kern 1pt} {\kern 1pt} {\kern 1pt} {\kern 1pt} {\kern 1pt} {\kern 1pt} {\kern 1pt} {\kern 1pt} {\kern 1pt} {\kern 1pt} {\kern 1pt} {\kern 1pt} {{R_k}(t) \ge R_k^{\min }(t)},\forall k, \forall t,\\
\left| {{\phi _n}(t)} \right| = 1,\forall n, \forall t,\\
{x_{\min }} \le x_{\text{UAV}}(t) \le {x_{\max }}, {y_{\min }} \le y_{\text{UAV}}(t) \le {y_{\max }}, \forall t,  \\
{R_{l,b \to l,a}} (t)  \ge {R_{l,b \to l,b}} (t), {\pi _l}(a) \ge {\pi _l}(b), \forall l, \\
{\text{tr}}\left( {{\boldsymbol{P}}{{\left( {{{\boldsymbol{H}}^H}{\boldsymbol{H}}} \right)}^{ - 1}}} \right) \le {P_{\max }} ,\forall k, \forall t,
\end{align}
\end{subequations}
\end{center}
where (21b) represents that the data demand of all MUs has to be satisfied at each timeslot, while $R_k^{\min }(t)$ denotes the minimal data rate requirement constraint for any MU. (21c) denotes that the passive beamforming constraint of the RIS need to be satisfied when controlling its phase shift. (21d) formulates the altitude bound of UAVs, which indicates that the UAV can only move in this particular area. (21e) is the decoding order constraint of NOMA technique for guaranteeing successful SIC. (21f) qualifies that the total required transmit power cannot exceed the maximal power constraint of the UAV. The formulated problem is a non-convex optimization problem. Additionally, the dynamic movement of the UAV and MUs are discussed, which indicates that the considered scenario is dynamic, which is challenging for the conventional optimization algorithms due to their failure to overcome the dynamic in the environment. Moreover, since joint trajectory design of the UAV, phase shift control of the RIS, decoding order and power allocation policy determination are planned to be optimized. The search-space is expanded as the number of parameters increases, which also makes the conventional gradient-based optimization techniques unsuitable. Therefore, RL algorithm, which empowers the agent to making decisions by learning from the environment, is invoked to solve the formulated problem.

\section{Proposed Solutions}

In this section, we first formulate the joint phase shift control and trajectory design problem as an MDP. Afterward, we proposed a D-DQN based algorithm for tackling the formulated problem. In addition, the state space, action space, reward function design of the proposed D-DQN based algorithm are specified.

\subsection{Markov Decision Process Formulation}

\begin{figure*}[t!]
    \begin{center}
        \includegraphics[width=14cm]{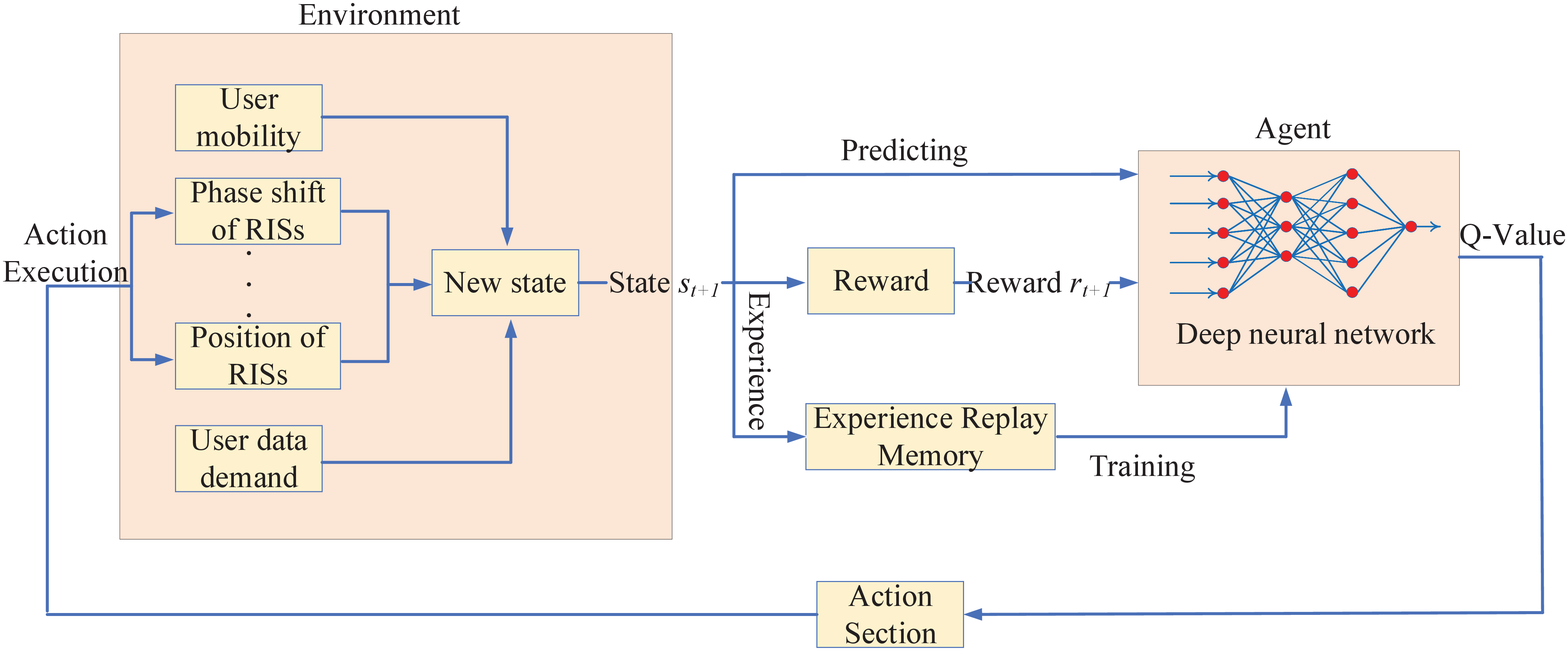}
     \caption{MDP and DRL model for RIS-aided UAv-enabled wireless network.}
            \label{DQN}
    \end{center}
\end{figure*}

Before invoking RL algorithms, the formulated problem is expected to be proved to be capable of being considered as an MDP problem. It has been proved in~\cite{zeng2020simultaneous} that, since the central controller makes sequential decisions, which influence the observed state (UAV's position, RIS's phase shifts, and allocated power to each MU) at next timeslot. Thus, the trajectory design and phase shift control problem for RIS-enhanced UAV-enabled wireless network can be formulated as an MDP. As illustrated in Fig.~\ref{DQN}, the MDP is defined by the environment, the set of states $S$, the set of actions $A$ and the reward function $r$ and the state transition function $\tau $ forms the MDP cycle. After one MDP cycle, the process transitions into a new state according to the previous state and the carried out actions.

\subsection{Proposed DRL Based Phase Shift Control and Trajectory Design Algorithm}

In this subsection, a D-DQN based algorithm is introduced to determine the trajectory of the UAV and the phase shift of the RIS, while guaranteeing that all the users' data demand is satisfied. In the D-DQN based model, the central controller, which controls both the UAV and the RIS, acts as the agent. At each time slot $t$, the agent observes a state, $s_t$, from the state space, $S$, which consists of the coordinates of both the UAV and of all the users, as well as of the phase shift of the RIS. According to the current state and decision policy $J$, the agent takes an action, $a_t$, from the action space, $A$, which consists of the moving directions of the UAV and the variable quantity of each reflecting element's phase shift. After carrying out actions, the agent obtains a reward/penalty $r_t$ based on the energy consumption of the UAV and the connectivity condition. At each timeslot, a Q-value is calculated based on the current state and previously taken actions. Thus, the state, action and Q-value is stored in a Q-function, $Q(s_t,a_t)$, which determines the decision policy $J$. The aim of the D-DQN model is to enable the agent to carry out the optimal actions to maximize the long-term sum reward. The principle of the D-DQN model is maximizing the long-term sum reward instead of aiming for maximizing the reward at a particular timeslot. Thus, in the D-DQN model, the selected action may not be the optimal choice for the current timeslot, but the optimal choice for pursing long-term benefits. In this paper, the phase shift of the RIS is considered as discrete, so value-based RL algorithm is invoked. When considering continuous phase shift, policy-based algorithms or actor-critic algorithms, such as DDPG algorithm can be adopted.

\begin{algorithm}[!t]
\caption{D-DQN based phase shift control and trajectory design algorithm}
\label{dq1}
\begin{algorithmic}
\REQUIRE ~~\\
Replay memory $D$, minibatch size $n$, and initial learning rate $\alpha$\\

\STATE \textbf{Initialize} the replay memory $D$, Q-network weights $\theta $, \\
 weights of the target network ${\theta ^*} = \theta $, and $Q(s,a)$.
\STATE \textbf{  } The UAV is deployed at a random point, the phase shift metric of the RIS is initially randomly decided.
\REPEAT
\STATE \textbf{For} each episode \textbf{do}:\\
\STATE  \textbf{  }\textbf{  }\textbf{  }\textbf{  }The central controller chooses $a_t$ uniformly with probability of $\varepsilon $, while chooses $a_t$ such that ${Q_\theta }({s_t},{a_t}) = {\max _{a \in A}}{Q_\theta }({s_t},{a_t})$ with probability of $(1-\varepsilon) $.
\STATE  \textbf{  }\textbf{  }\textbf{  }\textbf{  }The central controller observes reward $r_t$,
\STATE  \textbf{  }\textbf{  }\textbf{  }\textbf{  }The D-DQN model transfers to a new state ${s_{t+1}}$;
\STATE  \textbf{  }\textbf{  }\textbf{  }\textbf{  }Store transition $(s_t,a_t,r_t,{s_{t + 1}})$ and sample random
\STATE  minibatch of transitions ${(s_i,a_i,r_i,{{s'}_i})}_{i \in n}$ from $D$;
\STATE  \textbf{  }\textbf{  }\textbf{  }\textbf{  }\textbf{For} each $i \in I$, we can obtain
\STATE  \textbf{  }\textbf{  }\textbf{  }\textbf{  }\textbf{  }\textbf{  }\textbf{  }\textbf{  }\textbf{  }\textbf{  }\textbf{  }${y_i} = {r_i} + \gamma  \cdot {\max _{a \in A}}{Q_{{\theta ^*}}}({{s'}_i},a)$;
\STATE  \textbf{  }\textbf{  }\textbf{  }\textbf{  }Perform a gradient descent step
\STATE  \textbf{  }\textbf{  }\textbf{  }\textbf{  }$\theta  \leftarrow \theta  - {a_t} \cdot \frac{1}{I}\sum\limits_{i \in n} {[{y_i} - {Q_\theta }({s_i},{a_i})] \cdot {\nabla _\theta }{Q_\theta }({s_i},{a_i})} $;
\STATE  \textbf{  }\textbf{  }\textbf{  }\textbf{  }$\theta \leftarrow {\theta ^*}$.
\STATE  \textbf{  }\textbf{  }\textbf{  }\textbf{  }Calculate the learning rate based on $\alpha \left( {{n_e}} \right) = {{{\alpha _0}} \mathord{\left/
 {\vphantom {{{\alpha _0}} {1 + \eta {n_e}}}} \right.
 \kern-\nulldelimiterspace} {1 + \eta {n_e}}}$.
\STATE  \textbf{end}
\UNTIL{State $s$ terminates}
\ENSURE Q-function ${Q_\theta }$ and policy $J$.
\end{algorithmic}
\end{algorithm}

At each timeslot, the Q-value and Q-function are updated based on the current state, previously taken actions and the received reward by following the below principle

\begin{align}\label{Pout_n}
{\begin{gathered}
  {Q_{t + 1}}({s_t},{a_t}) \leftarrow (1 - \alpha ) \cdot {Q_t}({s_t},{a_t})  {\text{ + }}\alpha  \cdot \left[ {{r_t} + \gamma  \cdot {{\max }_a}{Q_t}({s_{t + 1}},a)} \right] ,\hfill \\
\end{gathered} }
\end{align}
where $\alpha $ denotes the learning rate and $\gamma $ is the discount factor.

In \eqref{Pout_n}, the reward $r_t$ is drawn from a fixed reward distribution $R:S\times A\to \mathbb{R}$, where $E\left\{ {{r}_{t}}\left| \left( s,a,{s}' \right)=\left( {{s}_{t}},{{a}_{t}},{{s}_{t+1}} \right) \right. \right\}=R_{sa}^{{{s}'}}$. By solving the following equation, the optimal value function can be obtained as
\begin{align}\label{Q*}
{ {{Q}^{*}}(s,a)={{\mathbb{E}}_{{{s}'}}}\left[ r+\gamma ma{{x}_{{{a}'}}}{{Q}^{*}}({s}',{a}')\left| s,a \right. \right] },
\end{align}
where ${Q}^{*}(s,a)$ is the optimal value function and $Q\to {{Q}^{*}}$ when $i\to \infty$.

\subsubsection{States in the D-DQN model}

In terms of the state space of the proposed D-DQN model. it contains four parts: 1) ${\theta _{n}}(t) \in \left[ {0,2\pi } \right],{\kern 1pt}{\kern 1pt}{\kern 1pt}n \in \mathcal{N}$, the phase shift metric of each reflecting elements at timeslot $t$; 2) $[x_{\text{UAV}}(t),y_{\text{UAV}}(t)]^T$, the 2D coordinate of the UAV at timeslot $t$\footnote[4]{In this paper, 2D trajectory design of the UAV is investigated due to lack of the energy consumption model of UAV's 3D movement. The results derived from the proposed algorithm can be extended to the 3D trajectory of the UAV.}; 3) ${c_k^U}(t) = {[{x_k^U}(t),{y_k^U}(t)]^T},{\kern 1pt}{\kern 1pt}{\kern 1pt}k \in \mathcal{K}$, the 2D coordinate of each MU at timeslot $t$; 4) $p_k(t),{\kern 1pt}{\kern 1pt}{\kern 1pt}k \in \mathcal{K}$, the power allocated from the UAV to each MU at timeslot $t$.

\subsubsection{Actions in the D-DQN model}

As for the action space of the proposed D-DQN model, it contains three parts: 1) $\Delta {\theta _{n}}(t)  \in  \left\{ { - \frac{\pi }{{10}},{\kern 1pt} {\kern 1pt} {\kern 1pt} 0,{\kern 1pt} {\kern 1pt} {\kern 1pt} \frac{\pi }{{10}}} \right\}$, the variable quantity of the phase shift value of each reflecting element; 2) $\Delta c_{\text{UAV}}^I(t) \in \left\{ {( - 1,0),{\kern 1pt} {\kern 1pt} {\kern 1pt} (1,0),{\kern 1pt} {\kern 1pt} {\kern 1pt} (0,0),{\kern 1pt} {\kern 1pt} {\kern 1pt} (0, - 1),{\kern 1pt} {\kern 1pt} {\kern 1pt} (0,1)} \right\}$, the traveling direction and distance of the UAV; 3) $\Delta {p_k}(t) \in \left\{ { - \widetilde p,0,\widetilde p} \right\}$, the variable quantity of the transmit power from the UAV to each MU.

\subsubsection{Reward function in the D-DQN model}

The reward/penalty function is decided by the transmit rate of each MU and the energy consumption of the UAV. Thus, the reward/penalty is a function of the UAV's coordinate, the RIS's phase shift metric, and the power allocation coefficient from the UAV to MUs, which can be calculated as $r(t)=f[x_{\text{UAV}}(t),y_{\text{UAV}}(t), {\theta _{n}}(t), {p_k}(t)]$. When the D-DQN model carries out an action that reduce the energy consumption while grantee the data transmit rate of each MU, a positive reward will be given to the agent. By taking any other actions, which result to the increment of the energy dissipation, unsatisfiable of the data transmit rate, the D-DQN model receives a penalty. Before designing the reward/penalty function of the D-DQN model, we invoke $\xi $ to value the satisfaction of users. Thus, we have

\begin{align}\label{Xi}
{\xi _k(t)} = \left\{ {\begin{array}{*{20}{c}}
  1&{{R_k}(t) \ge R_k^{\min },{\kern 1pt} {\kern 1pt} {\kern 1pt} \forall k \in \{ 1, \cdots ,K\} ,} \\
  0&{{R_k}(t) < R_k^{\min },{\kern 1pt} {\kern 1pt} {\kern 1pt} \forall k \in \{ 1, \cdots ,K\} ,}
\end{array}} \right.
\end{align}
where ${R_k^{\min }}$ denotes the minimal achievable rate of the $k$-th MU. Based on \eqref{Xi}, the reward/penalty function of the D-DQN model can be designed as

\begin{align}\label{reward}
{r(t) = \left\{ {\begin{array}{*{20}{l}}
  {C*\left( {\sum\nolimits_{k = 1}^K {{\xi _k}(t)}  - K} \right)*E(t + 1)}&{E(t + 1) > E(t){\text{\& }}\sum\nolimits_{k = 1}^K {{\xi _k}(t) < K} ,} \\
  {\left( {\sum\nolimits_{k = 1}^K {{\xi _k}(t)}  - K} \right)*E(t + 1)}&{E(t + 1) = E(t){\text{\& }}\sum\nolimits_{k = 1}^K {{\xi _k}(t) < K} ,} \\
  { - E(t + 1)}&{E(t + 1) > E(t){\text{\& }}\sum\nolimits_{k = 1}^K {{\xi _k}(t) = K} ,} \\
  {E(t + 1)}&{E(t + 1) = E(t){\text{\& }}\sum\nolimits_{k = 1}^K {{\xi _k}(t) = K} ,}
\end{array}} \right.}
\end{align}
where $C$ is of a constant value to guarantee that the penalty function for dissatisfaction of transmit rate is with a high value, so that actions that lead to this phenomenon can be avoided.

It can be observed from \eqref{reward} that, the UAV tends to carries out actions from controlling the phase shift of the RIS instead to changing the position of the UAV, unless the data rate of users cannot be satisfied. Additionally, maximizing the long-term sum rewards in the proposed D-DQN model tends to minimize the long-term energy consumption of the UAV.

Since the state space is huge, overflow happens when storing Q-value in the Q-table. In an effort to solve this problem, function approximation by neural networks is adopted for approximating Q-table. A Convolutional Neural Networks (CNN) with weights ${\rm{\{ }}\theta {\rm{\} }}$ is invoked to output the Q-table. In an effort to reduce the correlation of sampling, memory replay is invoked in the proposed D-DQN model. At the early stage of training, the agent carries out random implementation actions, and stores its experiences in a memory bank. The experiences, which contains the states, actions, and rewards, can be leveraged as training samples. The aim of CNN is to minimize the following loss function at each episode,

\begin{align}\label{Qfunction}
{{\rm{Loss}}(\theta ) = {\sum {\left[ {y - Q\left( {{s_t},{a_t},\theta } \right)} \right]} ^2}},
\end{align}
where we have $y = {r_t} + \gamma  \cdot \mathop {\max }\limits_{a \in A} {Q_{{\rm{old}}}}\left( {{s_t},{a_t},\theta } \right)$.

In terms of the number of neurons in hidden layers, this number has to be larger than the dimension of input metric and output metric. Thus, it depends on the antennas number of the UAV and reflecting elements number of the RIS, as well as the cluster number.

In order to strike a balance between exploration and exploitation in the proposed D-DQN algorithm, $\epsilon $-greedy exploration method is leveraged by satisfy the following principle

\begin{align}\label{sq7}
{\begin{gathered}
  Pr(J=\widehat{J}) =\left\{\begin{matrix}
1-\epsilon,  & \widehat{a}=\text{argmax}Q\left ( s,a \right ),
\\
\epsilon /\left ( \left | a \right |-1 \right ),& otherwise.
\end{matrix}\right.
\end{gathered}}
\end{align}

In terms of the learning rate, we invoke the decaying learning concept for attaining a tradeoff between accelerating training speed and converging to the local optimal, as well as for avoiding oscillation. The decaying learning rate is given by

\begin{align}\label{LR}
{ \alpha \left( {{n_e}} \right) = \frac{{{\alpha _0}}}{{1 + \eta {n_e}}}},
\end{align}
where ${{\alpha _0}}$ represents the learning rate at the initial episode, $\eta $ is a constant parameter for determining the decaying rate, ${{n_e}}$ denotes the training episodes.

\begin{remark}\label{remark:learning rate}
By invoking the decaying learning rate, the initial training episode is with a large learning rate, which is helpful for accelerating training speed. With the increasing of training episode, the learning rate decays, which is useful for the D-DQN model to converge to a local optimal.
\end{remark}

\subsection{Analysis of the Proposed Algorithm}

\subsubsection{Convergence analysis}

Before analyzing the convergence of the proposed D-DQN algorithm, the convergence of the conventional Q-learning algorithm and DQN algorithm has to be discussed first. Afterward, by discussing the influence of decaying learning rate on the convergence of the DQN algorithm, the convergence of the proposed D-DQN algorithm an be proved.

It has been proved in~\cite{melo2001convergence} that the conventional Q-learning algorithm converges to the optimal Q-function when satisfying $0 \le {\alpha _t} \le 1,{\kern 1pt} {\kern 1pt} {\kern 1pt} {\kern 1pt} \sum\limits_t {{\alpha _t} = \infty {\kern 1pt} } $ and $\sum\limits_t {\alpha _t^2 < \infty {\kern 1pt} } $. Additionally, it has also been proved that the DQN algorithm, which is an extended Q-learning algorithm, is capable of converging to the optimal state once the neural networks is large enough~\cite{Xiao2020Enhancing}. The DQN based algorithm is a sub-optimal solution due to the reason that the optimality of a RL model can not be guaranteed. In terms of the influence of decaying learning rate on the optimality and convergence of the DQN algorithm, since the only function of decaying learning rate concept is attaining a tradeoff between accelerating training speed and converging to the local optimal, as well as avoiding oscillation. Thus, the decaying learning rate concept will not affect the converging ability and optimality of the DQN algorithm, but the convergence rate will be influenced. Overall, the convergence of the proposed D-DQN algorithm can be guaranteed, while the convergence rate of the proposed D-DQN algorithm can also be influenced when comparing to the conventional DQN algorithm.

\subsubsection{Complexity analysis}

The computational complexity of the proposed D-DQN based algorithm consists of two aspects, namely, the computational complexity related to the CNN model and the computational complexity related to the training process. The computational complexity related to the CNN model can be calculated as $O\left[ {{f}_{1}}\left( n_{2}^{2}{{\left( {{n}_{1}}-{{n}_{2}}+1 \right)}^{2}}+{{f}_{2}}n_{3}^{2}{{\left( {{n}_{1}}-{{n}_{2}}-{{n}_{3}}+2 \right)}^{2}} \right) \right]$, where $i$ denotes the number of Conv layer~\cite{he2015convolutional}. The parameters in this equation is related to the number and size of the filters in each Conv layer. Since in the training stage, all states and actions are observed by the agent, the computational complexity related to the training process is given by $O(\left| S \right| \cdot \left| A \right|)$, where $\left| S \right|$ and $\left| A \right|$ are the total number of states and actions, respectively.

\section{Simulation Results}

In this section, we aim for verifying the validity of the proposed D-DQN based algorithm by illustrating the convergence of the proposed algorithm and for validating the enhancement of the network performance with the assistance of the RIS. Additionally, we test the performance of both NOMA-RIS case and OMA-RIS case. In the simulation, the UAV is initially placed at a random position and the phase shift metric of the RIS is also randomly given at initial timeslot. We invoke a 3 layer CNN with 50 nodes in hidden layer. The learning rate decays from 0.1 to 0.001. The simulation parameters are shown in Table~\ref{table:parameters}.

\begin{table}[htbp]\small
\caption{ Simulation parameters}
\centering
\begin{tabular}{|l||r||r|}
\hline
\centering
 Parameter & Description & Value\\
\hline
\centering
$f_c$ & Carrier frequency & 2GHz\\
 \hline
\centering
$V$ & Velocity of the UAV & 5m/s\\
 \hline
\centering
  $\alpha_{BM} $ & Path loss exponent for UAV-MU link & 3.5~\cite{zhang2019capacity,mu2019exploiting}\\
  \hline
  \centering
 $\alpha_{BS} $ & Path loss exponent for UAV-RIS link & 2.2~\cite{zhang2019capacity,mu2019exploiting}\\
  \hline
  \centering
 $\alpha_{SM} $ & Path loss exponent for RIS-MU link & 2.8~\cite{zhang2019capacity,mu2019exploiting}\\
  \hline
\centering
 $C_0$ & Path loss when $d_0=1$ & -30dB\\
 \hline
\centering
 $N_0$ & Noise power spectral density& -169dBm/Hz \\
 \hline
\centering
 $\alpha_0 $ & Initial learning rate & 0.1\\
 \hline
 \centering
  $\eta  $ & Decaying rate of the learning rate & 0.001\\
 \hline
 \centering
  $\varepsilon  $ & Exploration ratio & 0.1\\
 \hline
 \centering
 $\tau $ & Batch size & 128\\
 \hline
\centering
 $e $ & Replay memory size & 10000\\
 \hline
\centering
 $\beta $ & Discount factor & 0.7\\
 \hline
\end{tabular}
\label{table:parameters}
\end{table}

\begin{figure}
\centering
\includegraphics[width=3.5in]{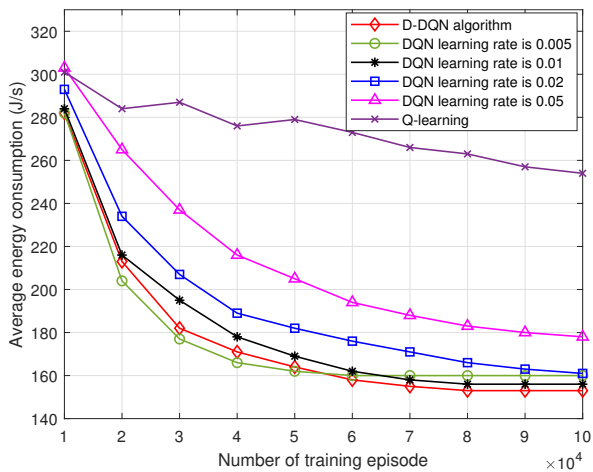}
\caption{Convergence rate of the proposed DQN algorithm.}\label{energyvsepisode}
\end{figure}

\textit{1) Convergence rate of the proposed D-DQN algorithm:} Fig.~\ref{energyvsepisode} characterizes the average energy consumption of the UAV over episodes. It can be observed that the conventional Q-learning algorithm fails to converge, which is due to the huge state space of UAV-aided wireless network. In contrast to the conventional Q-learning model, the D-DQN algorithm is capable of converging with the aid of the concept of function approximation via neural network. It can also be observed that when the learning rate is 0.005, the proposed DQN algorithm can converge after about 50000 episodes, which indicates that the performance of the DQN model (learning rate is 0.005) outperforms the cases with larger learning rate in terms of both converging rate and average energy consumption. Although the convergence rate of the proposed D-DQN based algorithm is slower than that of DQN algorithm (learning rate is 0.005), the average energy consumption derived from the proposed D-DQN algorithm is less than the DQN algorithm. The reason is that the decaying learning rate concept is helpful for attaining a tradeoff between accelerating training speed and converging to the local optimal.

\begin{figure}
\centering
\includegraphics[width=3.5in]{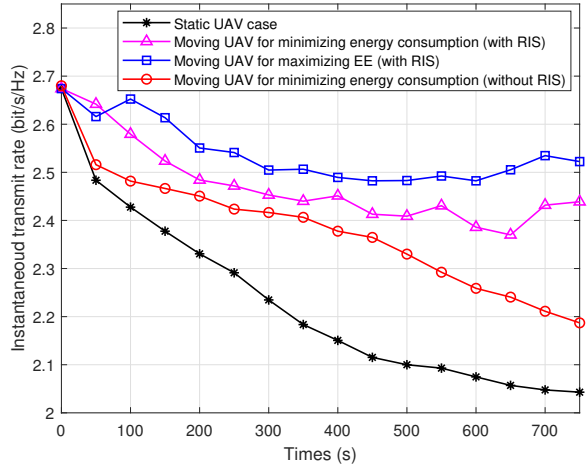}
\caption{Instantaneous transmit rate over time.}\label{timevsrate}
\end{figure}

\textit{2) Instantaneous transmit rate over time:} Fig.~\ref{timevsrate} characterizes the sum transmit rate at each timeslot. As mentioned before, the MUs are considered as roaming continuously, so the sum transmit rate of users are varying even the UAV is static. It can be observed from Fig.~\ref{timevsrate} that the instantaneous achievable rate of users decreases over time. The reason is that in the invoked walk model, users are with a roaming direction and velocity, which determines that users are walking away from their initial position. Thus, in the static UAV case, the distance between the UAV and users is increasing over time, which leads to a reduction of achievable rate. It can also be observed form Fig.~\ref{timevsrate} that by designing the trajectory of the UAV, the downtrend of achievable rate can be effectively slowed down. Additionally, Fig.~\ref{timevsrate} shows that with the aid of the RIS, a higher transmit rate can be achieved at each timeslot than the case of without RIS. It is worth noting that when the objective is to maximize the EE of the network, the achievable rate is higher than that of minimizing energy consumption. This is due to the reason that when targeting at minimizing energy consumption of the UAV, the UAV moves only when the data demand constraint of users is not satisfied, which means that the objective of this paper is not pursuing a higher throughput, but less movement of the UAV.

\begin{figure}
\centering
\includegraphics[width=3.5in]{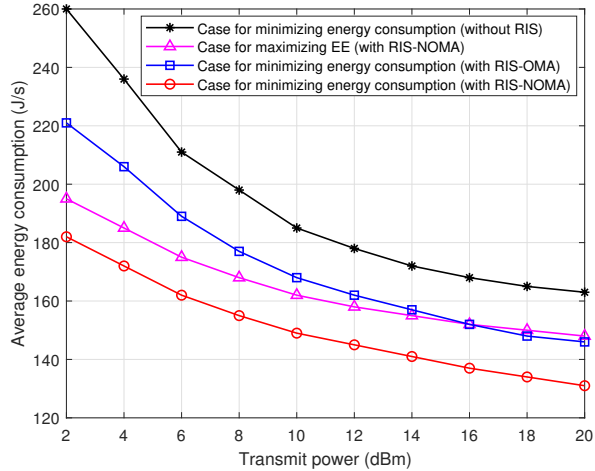}
\caption{Average energy consumption over transmit power. (The number of reflecting elements is 24)}\label{powervsenergy}
\end{figure}

\textit{3) Impact of transmit power:} Fig.~\ref{powervsenergy} characterizes the average energy consumption of the UAV over transmit power. It can be observed from Fig.~\ref{powervsenergy} that the UAV consumes less energy when increasing the transmit power. In the case of minimizing energy consumption without RIS, the UAV consumes the most energy. This is because that the UAV has to carry out more action of moving to form LoS link between it and MUs without the aid of the RIS. It can also be observed that even though a higher achievable rate can be obtained by maximizing EE, it will consume far more energy. Compared to the case of maximizing EE, the case of minimizing energy dissipation consume less energy while the data demand of all users is satisfied, which emphasizes the motivation of minimizing energy dissipation instead of maximizing EE. Finally, when comparing the RIS-NOMA case with the RIS-OMA case, we can observe that the RIS-NOMA case outperforms the RIS-OMA case in terms of energy consumption.

\begin{figure}
\centering
\includegraphics[width=3.5in]{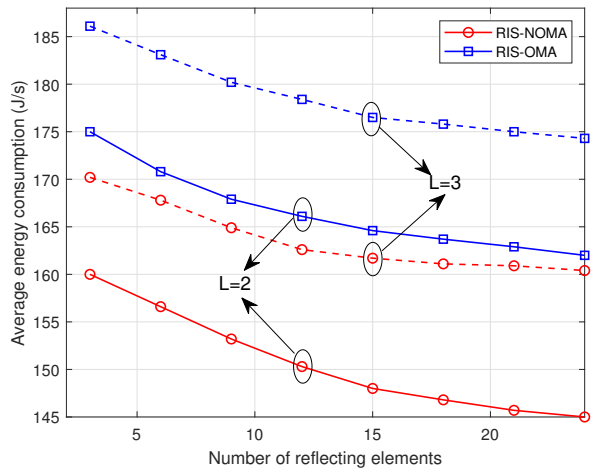}
\caption{Average energy consumption over number of reflecting elements. (The transmit power is 12dBm)}\label{numbervsenergy}
\end{figure}

\textit{4) Impact of reflecting elements number:} Fig.~\ref{numbervsenergy} characterizes the average energy consumption of the UAV over reflecting elements number. It shows in this figure that invoking more reflecting elements leads to the reduction of energy consumption. When serving more MUs, the UAV will consume more energy. For instance, in the case of cluster number is 3, the UAV will consume 10.3\% more energy than that of cluster number is 2. We can also observe that RIS-NOMA case consume 11.7\% less energy than RIS-OMA case. This is due to the reason that NOMA networks enjoy a higher spectrum efficiency than OMA networks. Since the transmit rate of MUs in NOMA networks is higher than that in OMA networks, the data demand constraint is more likely to be satisfied in NOMA networks, which indicates that less movement actions will be carried out by the UAV.

\begin{figure}
\centering
\includegraphics[width=3.5in]{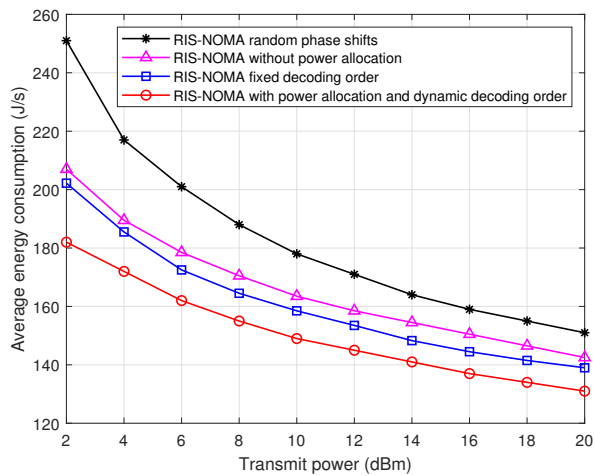}
\caption{Impact of decoding order and power allocation. (The number of reflecting elements is 24)}\label{orderallocation}
\end{figure}

\textit{5) Impact of decoding order and power allocation:} Fig.~\ref{orderallocation} characterizes the average energy consumption of the UAV in different cases. Since both the UAV and MUs are moving, the decoding order has to be re-determined at each timeslot to guarantee success SIC. When invoking a fixed decoding order, the spectrum efficiency of NOMA networks will decrease, which leads to the reduction of users' achievable rate. In order to satisfy the data demand constraint, the UAV has to move to provide high quality wireless services. Since NOMA technique superimposes the signals of two users at different powers, invoking a fixed power allocation policy leads to a higher energy consumption than attaining the optimal power allocation at each timeslot. Thus, Fig.~\ref{orderallocation} shows that by determining the dynamic decoding order and power allocation policy, the energy consumption can be reduced. It can also be observed that by designing the phase shift of the RIS, the energy consumption can be significantly reduced compared to random phase shift case.

\begin{figure}
\centering
\includegraphics[width=3.5in]{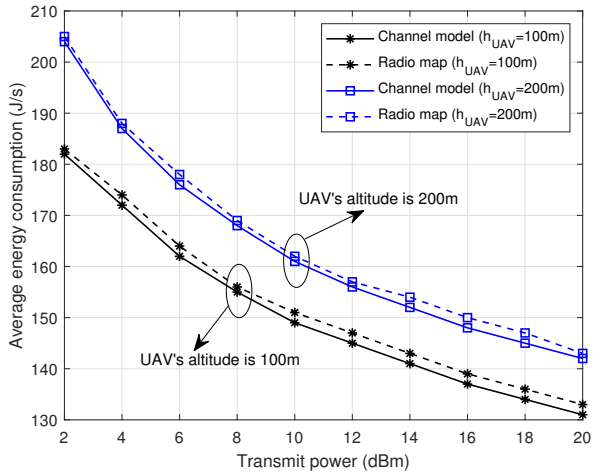}
\caption{Impact of UAV's altitude on energy consumption. (The number of reflecting elements is 24)}\label{energyvsaltitude}
\end{figure}

\textit{6) Impact of UAV's altitude:} Fig.~\ref{energyvsaltitude} characterizes the average energy consumption of the UAV at different altitude. It can be observed that when the altitude of the UAV is 200m, it consumes more energy. This is due to the reason that increasing the altitude of the UAV increase the LoS probability between the UAV and MUs but leads to a higher path loss due to the augment of distance. We can also observe that the result derived from invoking 3GPP channel model is not exactly the same as that derived from invoking 3D radio map. The reason is that the buildings in the real world is with irregularly-shape, but buildings are constructed by cubes in the radio map. However, the variation trends of two curves are the same, which proves the validity of the proposed approaches.

\section{Conclusions}

In this paper, a novel framework for leveraging RISs and NOMA technique in UAV-enabled wireless network has been proposed. The problem of jointly phase shift control of the RIS, dynamic trajectory design of the UAV, dynamic decoding order determination in NOMA technique, and power allocation policy determination has been formulated for minimizing the energy consumption of the UAV by considering its propulsion-related energy consumption. Additionally, the data demand of each MUs has been satisfied at each timeslot. A D-DQN based algorithm, which is the combination of DQN algorithm and decaying learning rate concept, has been proposed for solving the formulated algorithm. The proposed D-DQN based algorithm has been proved to be capable of striking a balance between accelerating training speed and converging to the local optimal, as well as for avoiding oscillation. It has been proved by simulation results that, with the aid of the RIS, the energy consumption of the UAV can be significantly reduced while the RIS-NOMA case consumes less energy than the RIS-OMA case. Our future work will optimize the velocity of the UAV for further enhancing the performance of the RIS-UAV enabled wireless networks.

\begin{spacing}{1.45}
\bibliographystyle{IEEEtran}
\bibliography{mybib}

\begin{thebibliography}{10}
\providecommand{\url}[1]{#1}
\csname url@samestyle\endcsname
\providecommand{\newblock}{\relax}
\providecommand{\bibinfo}[2]{#2}
\providecommand{\BIBentrySTDinterwordspacing}{\spaceskip=0pt\relax}
\providecommand{\BIBentryALTinterwordstretchfactor}{4}
\providecommand{\BIBentryALTinterwordspacing}{\spaceskip=\fontdimen2\font plus
\BIBentryALTinterwordstretchfactor\fontdimen3\font minus
  \fontdimen4\font\relax}
\providecommand{\BIBforeignlanguage}[2]{{%
\expandafter\ifx\csname l@#1\endcsname\relax
\typeout{** WARNING: IEEEtran.bst: No hyphenation pattern has been}%
\typeout{** loaded for the language `#1'. Using the pattern for}%
\typeout{** the default language instead.}%
\else
\language=\csname l@#1\endcsname
\fi
#2}}
\providecommand{\BIBdecl}{\relax}
\BIBdecl

\bibitem{liu2020Reconfigurable}
Y.~Liu, X.~Liu, X.~Mu, T.~Hou, J.~Xu, Z.~Qin, M.~Di~Renzo, and N.~Al-Dhahir,
  ``Reconfigurable intelligent surfaces: Principles and opportunities,''
  \emph{arXiv:2007.03435}, 2020.

\bibitem{Qingqing2020Towards}
Q.~{Wu} and R.~{Zhang}, ``Towards smart and reconfigurable environment:
  Intelligent reflecting surface aided wireless network,'' \emph{{IEEE} Commun.
  Mag.}, vol.~58, no.~1, pp. 106--112, 2020.

\bibitem{najafi2020physics}
M.~Najafi, V.~Jamali, R.~Schober, and V.~H. Poor, ``Physics-based modeling and
  scalable optimization of large intelligent reflecting surfaces,''
  \emph{arXiv:2004.12957}, 2020.

\bibitem{liu2019trajectory}
X.~Liu, Y.~Liu, Y.~Chen, and L.~Hanzo, ``Trajectory design and power control
  for multi-{UAV} assisted wireless networks: A machine learning approach,''
  \emph{{IEEE} Trans. Veh. Technol.}, vol.~68, no.~8, pp. 7957 -- 7969, 2019.

\bibitem{yan2019comprehensive}
C.~Yan, L.~Fu, J.~Zhang, and J.~Wang, ``A comprehensive survey on {UAV}
  communication channel modeling,'' \emph{IEEE Access}, vol.~7, pp.
  107\,769--107\,792, 2019.

\bibitem{mozaffari2019tutorial}
M.~Mozaffari, W.~Saad, M.~Bennis, Y.-H. Nam, and M.~Debbah, ``A tutorial on
  {UAVs} for wireless networks: Applications, challenges, and open problems,''
  \emph{{IEEE} Commun. Surveys Tutorials}, vol.~21, no.~3, pp. 2334--2360,
  2019.

\bibitem{wu2018joint}
Q.~Wu, Y.~Zeng, and R.~Zhang, ``Joint trajectory and communication design for
  multi-{UAV} enabled wireless networks,'' \emph{{IEEE} Trans. Wireless
  Commun.}, vol.~17, no.~3, pp. 2109--2121, 2018.

\bibitem{mozaffari2017mobile}
M.~Mozaffari, W.~Saad, M.~Bennis, and M.~Debbah, ``Mobile unmanned aerial
  vehicles ({UAVs}) for energy-efficient internet of things communications,''
  \emph{{IEEE} Trans. Wireless Commun.}, vol.~16, no.~11, pp. 7574--7589, 2017.

\bibitem{liu2018energy}
C.~H. Liu, Z.~Chen, J.~Tang, J.~Xu, and C.~Piao, ``Energy-efficient {UAV}
  control for effective and fair communication coverage: A deep reinforcement
  learning approach,'' \emph{{IEEE} J. Sel. Areas Commun.}, vol.~36, no.~9, pp.
  2059--2070, 2018.

\bibitem{cui2019multi}
J.~Cui, Y.~Liu, and A.~Nallanathan, ``Multi-agent reinforcement learning-based
  resource allocation for {UAV} networks,'' \emph{{IEEE} Trans. Wireless
  Commun.}, vol.~19, no.~2, pp. 729--743, 2019.

\bibitem{liu2019distributed}
C.~H. Liu, X.~Ma, X.~Gao, and J.~Tang, ``Distributed energy-efficient
  multi-{UAV} navigation for long-term communication coverage by deep
  reinforcement learning,'' \emph{{IEEE} Trans. Mobile Computing.}, accept to
  appear, 2019.

\bibitem{guo2019weighted}
H.~Guo, Y.-C. Liang, J.~Chen, and E.~G. Larsson, ``Weighted sum-rate
  optimization for intelligent reflecting surface enhanced wireless networks,''
  \emph{arXiv:1905.07920}, 2019.

\bibitem{wu2019intelligent}
Q.~Wu and R.~Zhang, ``Intelligent reflecting surface enhanced wireless network
  via joint active and passive beamforming,'' \emph{{IEEE} Trans. Wireless
  Commun.}, vol.~18, no.~11, pp. 5394--5409, 2019.

\bibitem{huang2019reconfigurable}
C.~Huang, A.~Zappone, G.~C. Alexandropoulos, M.~Debbah, and C.~Yuen,
  ``Reconfigurable intelligent surfaces for energy efficiency in wireless
  communication,'' \emph{{IEEE} Trans. Wireless Commun.}, 2019.

\bibitem{shen2019secrecy}
H.~Shen, W.~Xu, S.~Gong, Z.~He, and C.~Zhao, ``Secrecy rate maximization for
  intelligent reflecting surface assisted multi-antenna communications,''
  \emph{{IEEE} Commun. Lett.}, vol.~23, no.~9, pp. 1488--1492, 2019.

\bibitem{huang2020reconfigurable}
C.~Huang, R.~Mo, C.~Yuen \emph{et~al.}, ``Reconfigurable intelligent surface
  assisted multiuser miso systems exploiting deep reinforcement learning,''
  \emph{arXiv:2002.10072}, 2020.

\bibitem{feng2020deep}
K.~Feng, Q.~Wang, X.~Li, and C.-K. Wen, ``Deep reinforcement learning based
  intelligent reflecting surface optimization for {MISO} communication
  systems,'' \emph{{IEEE} Wireless Commun. Lett.}, 2020.

\bibitem{taha2020deep}
A.~Taha, Y.~Zhang, F.~B. Mismar, and A.~Alkhateeb, ``Deep reinforcement
  learning for intelligent reflecting surfaces: Towards standalone operation,''
  \emph{arXiv:2002.11101}, 2020.

\bibitem{Helin2020Deep}
J.~Z. D.~N. Helin~Yang, Zehui~Xiong and L.~Xiao, ``Deep reinforcement learning
  based intelligent reflecting surface for secure wireless communications,''
  \emph{arXiv:2002.12271}, 2020.

\bibitem{zhang2020millimeter}
Q.~Zhang, W.~Saad, and M.~Bennis, ``Millimeter wave communications with an
  intelligent reflector: Performance optimization and distributional
  reinforcement learning,'' \emph{arXiv:2002.10572}, 2020.

\bibitem{liu2017non}
Y.~Liu, Z.~Qin, M.~Elkashlan, Z.~Ding, A.~Nallanathan, and L.~Hanzo,
  ``Non-orthogonal multiple access for {5G} and beyond,'' vol. 105, no.~12, pp.
  2347--2381, 2017.

\bibitem{zhang2020energy}
H.~Zhang, J.~Zhang, and K.~Long, ``Energy efficiency optimization for {NOMA}
  {UAV} network with imperfect {CSI},'' \emph{arXiv:2005.02046}, 2020.

\bibitem{lu2020uav}
J.~Lu, Y.~Wang, T.~Liu, Z.~Zhuang, X.~Zhou, F.~Shu, and Z.~Han, ``{UAV}-enabled
  uplink non-orthogonal multiple access system: Joint deployment and power
  control,'' \emph{{IEEE} Trans. Veh. Technol.}, accept to appear, 2020.

\bibitem{liu2020ris}
X.~Liu, Y.~Liu, Y.~Chen, and H.~V. Poor, ``{RIS} enhanced massive
  non-orthogonal multiple access networks: Deployment and passive beamforming
  design,'' \emph{arXiv:2001.10363}, 2020.

\bibitem{mu2019exploiting}
X.~Mu, Y.~Liu, L.~Guo, J.~Lin, and N.~Al-Dhahir, ``Exploiting intelligent
  reflecting surfaces in multi-antenna aided {NOMA} systems,''
  \emph{arXiv:1910.13636}, 2019.

\bibitem{di2019smart}
M.~Di~Renzo, M.~Debbah, D.-T. Phan-Huy, A.~Zappone, M.-S. Alouini, C.~Yuen,
  V.~Sciancalepore, G.~C. Alexandropoulos, J.~Hoydis, H.~Gacanin \emph{et~al.},
  ``Smart radio environments empowered by {AI} reconfigurable meta-surfaces: An
  idea whose time has come,'' \emph{arXiv:1903.08925}, 2019.

\bibitem{qingqing2019towards}
Q.~{Wu} and R.~{Zhang}, ``Towards smart and reconfigurable environment:
  Intelligent reflecting surface aided wireless network,'' \emph{{IEEE} Commun.
  Mag.}, vol.~58, no.~1, pp. 106--112, 2020.

\bibitem{wu2018intelligent}
Q.~Wu and R.~Zhang, ``Intelligent reflecting surface enhanced wireless network
  via joint active and passive beamforming,'' \emph{{IEEE} Trans. Wireless
  Commun.}, 2019.

\bibitem{abeywickrama2019intelligent}
S.~Abeywickrama, R.~Zhang, and C.~Yuen, ``Intelligent reflecting surface:
  Practical phase shift model and beamforming optimization,'' \emph{{IEEE}
  Trans. Commun.}

\bibitem{muruganathan2018overview}
S.~D. Muruganathan, X.~Lin, H.-L. Maattanen, Z.~Zou, W.~A. Hapsari, and
  S.~Yasukawa, ``An overview of {3GPP} release-15 study on enhanced {LTE}
  support for connected drones,'' \emph{arXiv preprint arXiv:1805.00826}, 2018.

\bibitem{zeng2019energy}
Y.~Zeng, J.~Xu, and R.~Zhang, ``Energy minimization for wireless communication
  with rotary-wing {UAV},'' \emph{{IEEE} Trans. Wireless Commun.}, vol.~18,
  no.~4, pp. 2329--2345, 2019.

\bibitem{liu2016cooperative}
Y.~Liu, Z.~Ding, M.~Elkashlan, and H.~V. Poor, ``Cooperative non-orthogonal
  multiple access with simultaneous wireless information and power transfer,''
  \emph{{IEEE} J. Sel. Areas Commun.}, vol.~34, no.~4, pp. 938--953, 2016.

\bibitem{ding2017survey}
Z.~Ding, X.~Lei, G.~K. Karagiannidis, R.~Schober, J.~Yuan, and V.~K. Bhargava,
  ``A survey on non-orthogonal multiple access for {5G} networks: Research
  challenges and future trends,'' \emph{{IEEE} J. Sel. Areas Commun.}, vol.~35,
  no.~10, pp. 2181--2195, 2017.

\bibitem{liu2018multiple}
Y.~Liu, H.~Xing, C.~Pan, A.~Nallanathan, M.~Elkashlan, and L.~Hanzo,
  ``Multiple-antenna-assisted non-orthogonal multiple access,'' \emph{{IEEE}
  Wireless Commun.}, vol.~25, no.~2, pp. 17--23, 2018.

\bibitem{basar2019large}
E.~Basar, ``Large intelligent surface-based index modulation: A new beyond
  {MIMO} paradigm for 6{G},'' \emph{arXiv:1904.06704}, 2019.

\bibitem{huang2018energy}
C.~Huang, G.~C. Alexandropoulos, A.~Zappone, M.~Debbah, and C.~Yuen, ``Energy
  efficient multi-user {MISO} communication using low resolution large
  intelligent surfaces,'' in \emph{IEEE Proc. of Global Commun. Conf.
  (GLOBECOM)}, 2018, pp. 1--6.

\bibitem{cui2017optimal}
J.~Cui, Y.~Liu, Z.~Ding, P.~Fan, and A.~Nallanatha, ``Optimal user scheduling
  and power allocation for millimeter wave {NOMA} systems,'' \emph{{IEEE}
  Trans. Wireless Commun.}, vol.~17, no.~3, pp. 1502--1517, 2017.

\bibitem{zeng2020simultaneous}
Y.~Zeng, X.~Xu, S.~Jin, and R.~Zhang, ``Simultaneous navigation and radio
  mapping for cellular-connected {UAV} with deep reinforcement learning,''
  \emph{arXiv:2003.07574}, 2020.

\bibitem{melo2001convergence}
F.~S. Melo, ``Convergence of {Q-learning}: {A} simple proof,'' \emph{Institute
  Of Systems and Robotics, Tech. Rep}, pp. 1--4, 2001.

\bibitem{Xiao2020Enhancing}
X.~{Liu}, Y.~{Liu}, Y.~{Chen}, and L.~{Hanzo}, ``Enhancing the fuel-economy of
  v2i-assisted autonomous driving: A reinforcement learning approach,''
  \emph{{IEEE} Trans. Veh. Technol.}, acccept to appear, 2020.

\bibitem{he2015convolutional}
K.~He and J.~Sun, ``Convolutional neural networks at constrained time cost,''
  in \emph{Proceedings of the IEEE conference on computer vision and pattern
  recognition}, 2015, pp. 5353--5360.

\bibitem{zhang2019capacity}
S.~Zhang and R.~Zhang, ``Capacity characterization for intelligent reflecting
  surface aided {MIMO} communication,'' \emph{{IEEE} J. Sel. Areas Commun.},
  accept to appear, 2020.

\end{thebibliography}
\end{spacing}
\end{document}